\documentclass[a4paper,UKenglish]{lipicsown}

\def\spaceo{\setlength\itemsep{0.5em}}
\def\fracy#1#2{{#1}/{#2}}
\usepackage{amsfonts}
\usepackage{amssymb}
\usepackage{float}
\usepackage{enumerate}
\usepackage{tikz}
\newtheorem{exmp}[theorem]{Example}
\usepackage{amsmath}
\usepackage{mathtools}

\newlength\myindent 
\title{Improved Algorithms for the General Exact Satisfiability Problem}
\titlerunning{Improved Algorithms for the General Exact Satisfiability Problem}

\author[1]{Gordon Hoi}
\affil[1]{School of Computing, National University of Singapore,
13 Computing Drive, Block COM1, Singapore 117417,
Republic of Singapore, \texttt{e0013185@u.nus.edu};
Since February 2021 G.~Hoi was
supported by the Singapore Ministry of Education Academic
Research Fund Tier 2 grant MOE2019-T2-2-121 / R146-000-304-112.}
\author[2]{Frank Stephan}
\affil[2]{Department of Mathematics and School of
Computing, National University of Singapore,
10 Lower Kent Ridge Road, Block S17, Singapore 119076,
Republic of Singapore, \texttt{fstephan@comp.nus.edu.sg};
F.~Stephan was supported part by the Singapore Ministry of Education Academic
Research Fund Tier 2 grant MOE2019-T2-2-121 / R146-000-304-112.}

\authorrunning{G.~Hoi and F.~Stephan}

\date{\today}

\begin{document}

\maketitle

\begin{abstract}
\noindent
The Exact Satisfiability problem asks if we can find a satisfying assignment
to each clause such that exactly one literal in each clause is assigned $1$,
while the rest are all assigned $0$. We can generalise this problem
further by defining that a $C^j$ clause is solved iff
exactly $j$ of the literals in the clause are $1$ and all others are $0$.
We now introduce the family of Generalised Exact Satisfiability
problems called G$i$XSAT as the problem to check whether a given instance
consisting of $C^j$ clauses with $j \in \{0,1,\ldots,i\}$ for each
clause has a satisfying assignment. In this paper, we
present faster exact polynomial space algorithms, using a nonstandard measure, 
to solve G$i$XSAT, for $i\in \{2,3,4\}$,
in $O(1.3674^n)$ time, $O(1.5687^n)$ time and $O(1.6545^n)$ time,
respectively, using polynomial space, where $n$ is the number
of variables. This improves the current state
of the art for polynomial space algorithms from $O(1.4203^n)$ time
for G$2$XSAT by Zhou, Jiang and Yin and from
$O(1.6202^n)$ time for G$3$XSAT by Dahll\"of and from $O(1.6844^n)$ time for
G$4$XSAT which was by Dahll\"of as well.
In addition, we present faster exact algorithms solving G$2$XSAT, G$3$XSAT
and G$4$XSAT in $O(1.3188^n)$ time, $O(1.3407^n)$ time and $O(1.3536^n)$ time
respectively at the expense of using exponential space.
\end{abstract}

\section{Introduction}  

\noindent
The Satisfiability problem, SAT, is an important problem in
computational complexity theory because it has been commonly 
used as a framework to solve other combinatorial
problems and it is known to be NP-complete \cite{Cook71}.
In addition, SAT has found many uses in practice as well. Applications
include AI-planning and software model checking \cite{Silva08,NOT06}.

Because of its importance, many variants has also been explored and studied.
One such variant is the Exact Satisfiability problem, XSAT,
where it asks if one can find a satisfying assignment such that exactly one
literal in each clause is assigned $1$ (true),
while the rest are assigned $0$ (false). XSAT is also known as $1$-in-SAT. 
This problem has been widely studied; Dahll\"of \cite{D06} provided an
algorithm running in $O(1.1730^n)$; a prior algorithm of Byskov, Madsen
and Skjernaa was running in $O(1.1749^n)$. Recent work of Hoi improved the
bound to $O(1.1674^n)$ \cite{H20}.
One may generalise the XSAT problem by considering
a constant bound $i$ and allowing an arbitrary collection of $C^j$
clauses as an instance, $j \leq i$, where each $C^j$ clause must
have exactly $j$ literals true for being satisfied.
This class of problems is also known as the General Exact
Satisfiability problem and we distinguish each individual 
problem as G$i$XSAT (General $i$ Exact Satisfiability) with
the bound $i$ as just described; the problem is also known
as $i$-in-SAT or X$_i$SAT. G$i$XSAT is of interest because
one can use it to represent cardinality constraints in propositional
logic modelling. G$i$XSAT sits in the middle of XSAT and Knapsack.
The implication XSAT to G$i$XSAT is immediate by the definition
and the standard reduction of XSAT to the Knapsack problem can easily
be modified to a reduction of the G$i$XXSAT problem to the Knapsack problem
where the number of variables does not increase.
Note that there is no restriction on the length of
the clauses here. Therefore, XSAT is also equivalent 
to G$1$XSAT. For any $i\geq1$, G$i$XSAT is known to be
NP-complete \cite{D05,D06,Sch78}. 

In this paper, we focus on G$i$XSAT, where $i \in \{2,3,4\}$. Dahll\"of
gave the first polynomial space algorithm solving G$2$XSAT running
in $O(1.5157^n)$ time. In his same paper \cite{D05}, he improved the
bound to $O(1.4511^n)$ time. In addition,
he also gave an algorithm to solve G$3$XSAT and G$4$XSAT in $O(1.6202^n)$ time
and $O(1.6844^n)$ time respectively.  
More recently, Zhou, Jiang and Yin \cite{ZJY14} gave an algorithm to solve
G$2$XSAT in $O(1.4203^n)$ time, improving upon Dahll\"of's G$2$XSAT algorithm.
These algorithms all use polynomial space. 

In his paper, Dahll\"of also gave an exact algorithms to solve G$2$XSAT
in $O(1.4143^n)$ time, at the expense of using $O(1.1893^n)$ space.
His result is a direct application of the Split and List technique
by Schroeppel and Shamir \cite{SS81}.

In this paper, we propose an algorithm to solve G$2$XSAT, G$3$XSAT and G$4$XSAT
in time $O(1.3674^n)$ time, $O(1.5687^n)$ time and in $O(1.6545^n)$ time 
respectively. $C^{j+1}$ clauses are more complex than $C^j$ clauses as
there are many more ways to satisfy them, the complexity of G$(i+1)$XSAT
is expected to be larger than that of G$i$XSAT and the running time
of the so far published algorithms reflect this. The improved bounds were
obtained by more comprehensive case-distinctions (in the case of G$2$XSAT)
as well as the use of a nonstandard measure for all three variants.
Furthermore, for G$2$XSAT, we use nonstandard measure in
the form of a state based measure to bring down the timing further.

In addition, we propose faster exact algorithms to solve G$2$XSAT, G$3$XSAT
and G$4$XSAT in time $O(1.3188^n)$ time, $O(1.3407^n)$ time
and in $O(1.3536^n)$ time, respectively,
using exponential space. To achieve this, we modify the technique
of Schroeppel and Shamir by doing asymmetric splits of the set of
variables such that the larger one of the two ``halves'' is a union
of clauses and there we exploit that for each $C^j$ clause of $\ell$ literals
not all $2^\ell$ possible values of the literals apply but only
${\ell \choose j}$ many and this effect brings down the overall number
of assignments for this larger half; however, one cannot do this for both
halves, as one cannot guarantee that the smaller half is fully covered by
clauses using only variables from that half.

\section{Preliminaries}

\subsection{Branching factor and vector}

\noindent
In this section, we will introduce some definitions that we will use repeatedly in this paper and also the techniques needed
to understand the analysis of the algorithm. The algorithm that we design is a Davis–Putnam–Logemann–Loveland (DPLL) 
\cite{DLL60,DP60} style algorithm, or also known as branch and bound or branch and reduce algorithms.
DPLL algorithms are recursive in nature and have two kinds of rules associated with them : Simplification and Branching rules.
Simplification rules help us to simplify a problem instance or to act as a case to terminate the algorithm.
Branching rules on the other hand, help us to solve a problem instance by recursively solving smaller instances of the problem. 
To help us to better understand the execution of a DPLL algorithm, the notion of a search tree is commonly used. 
We can assign the root node of the search tree to be the original problem, while subsequent child nodes are assigned 
to be the smaller instances of the problem whenever we invoke a branching rule. For more information of this area,
one may refer to the textbook written by Fomin and Kratsch \cite{FK10}.

In order for us to analyse the running time of the DPLL algorithm, one may just bound the number of leaves generated
by the search tree. We let $\mu$ be the choice of our measure. Then we let $T(\mu)$ denote the maximum number of
leaves based on the measure $\mu$ on a branching rule. Therefore, in order to know the worst case complexity
of running the algorithm, we need to analyse each branching rule separately and use the worst case runtime over all
the branching rules as an upper bound of the algorithm.

Now given any branching rule, let $r\geq2$ be the number of instances generated from this rule. Let
$t_1,t_2 ,...,t_r$ be the change of measure for each instance for a branching rule, then we have that
$T(\mu) \leq T(\mu-t_1) + T(\mu-t_2) + ... T(\mu-t_r)$, which is a linear recurrence. We can employ 
techniques in \cite{Kul99} to solve it. The number of leaves generated of this branching rule is therefore 
given as $\beta$, where $\beta$ is the unique positive root of $x^{-t_1} + x^{-t_2} + ... + x^{-t_r} = 1$. 
For ease of writing, we denote the branching factor of this branching rule as $\tau(t_1,t_2,...,t_r)=\beta$ and
 $(t_1,t_2,...t_r)$ is also known as the branching vector. 

If there are $k$ branching rules in the entire DPLL algorithm, with each having a branching factor of
$\beta_1,\beta_2,...,\beta_k$, then the entire algorithm runs in $O(c^{\mu})$, where
$c=max\{\beta_1,\beta_2,...,\beta_k\}$. 

We will introduce some known results about branching factors. If $k \leq k'$
and $j \leq j'$ then we have that
$\tau(k',j') \leq \tau(k,j)$, for all positive $k,j$.
In other words, comparing two branching factors,
if in each branch the first one eliminates at least as much
weight as the second, then the first one is as a number better or equal (that
is, as a number less or equal) than the second branching factor. The same
is true for multiple-way branching provided that the number of entries
for both branching factors is the same.
Suppose that 
$i+j = 2k$, for some $i,j,k$, then $\tau(k,k) \leq \tau(i,j)$. In other words, 
a more balanced tree will result in a smaller branching factor. 

Finally, suppose that we have a branching vector of
$(u,v)$ for some branching rule. Suppose
that for the first branch, we immediately do a follow up branching
to get a branching vector 
of $(t,w)$, then we can apply branching vector addition
to get a combined branching 
vector of $(u+t,u+w,v)$. This technique can sometimes help
us to bring down the 
overall complexity of the algorithm further.

Finally, correctness of DPLL algorithms usually follows from the fact
that all cases have been covered. 
Now, we give a few definitions before moving onto the actual algorithm. 

\subsection{Definitions}

\begin{definition}
A clause $C$ is a disjunction of literals. Alternatively,
we also say that a clause $C$ is a multiset of literals. 
A $k$-literal clause $C$ is a clause where $|C|=k$.
Finally, $\alpha$ is a subclause of $C$ if $\alpha \subset C$. 
\end{definition}

\noindent
For example, if $C=(a \vee b \vee c \vee d)$, then $C=\{a,b,c,d\}$ is a 4-literal clause. In addition, let
$\delta = (a \vee b \vee c)$. Then $\delta$ is a subclause of $C$ since $\delta \subset C$ and we can write 
$C=(d \vee \delta)$. We will use greek letters for subclauses.

\begin{definition}
Two clauses $C_1$ and $C_2$ are called neighbouring clauses if they share a common variable. 
Let $Var(C)$ denote the set of variables that appear in the clause $C$.
In other words, $C_1$ and $C_2$ are neighbours when $Var(C_1) \cap Var(C_2) \neq \emptyset$. 
Let two clauses $C_1$ and $C_2$ be given such that $|Var(C_1) \cap Var(C_2)| = k\geq2$. We say that
they $C_1$ and $C_2$ have $k\geq2$ overlapping variables. 
\end{definition}

\noindent
For example, if we have a clause $C=(x \vee x \vee \neg y \vee z)$, then $Var(C)=\{x,y,z\}$.

\begin{definition}
We call a clause $C^i$ if it needs exactly $i$ literal(s) to be assigned true. Likewise,
we say that a clause is exact-$i$-satisfiable if we can assign $i$ literals $1$,
and the rest $0$ in that clause. For simplicity, we will just say that the clause is satisfiable
instead of exact-$i$-satisfiable. 
\end{definition}

\noindent
In the midst of branching a $C^j$ clause, it can drop to a $C^i$ clause, where
$i<j$.
Therefore, we introduce this definition to distinguish the different types
of clauses present in the formula.

\begin{definition}
We say that two variables, $x$ and $y$, are linked when we can deduce either $x=y$ or $x=\neg y$. When this happens,
we can proceed to remove one of the linked variables, either $x$ or $y$.
\end{definition}

\noindent
For example, if we have a $C^1$ clause $(x \vee y \vee z)$ and if $x=0$, we can deduce that $y =\neg z$. 
Then, we can replace all occurrences of $y$ by $\neg z$ and $\neg y$ by $z$.

\begin{definition}
Let $deg(x)$ denote the degree of a variable $x$, which is the number of times that the literal $x$ and $\neg x$ appears in 
the formula. We call a variable heavy if $deg(x)\geq3$. 
\end{definition}

\noindent
This definition will be mainly used in our G$2$XSAT (polynomial space) algorithm where we have to deal with 
heavy variables in $C^2$ clauses. Let $C_1 = (x \vee a_1 \vee a_2 \vee a_3)$, 
$C_2 = (x \vee a_4 \vee a_5 \vee a_6)$ and $C_3 = (x \vee a_7 \vee a_8 \vee a_9)$. Since $deg(x)\geq3$,
$x$ is heavy in this case.

\begin{definition}
Let $\varphi$ be a formula given.
Then we denote $\varphi[x=1]$ ($\varphi[x=0]$) as the formula obtained 
after assigning $x=1$ ($x=0$). Likewise,
we denote $\varphi[x=y]$ be the formula obtained after
replacing all occurrences of $x$ by $y$. Let $\alpha$ be a subclause of
some clause $C$. We denote $\varphi[\alpha=i]$ as the formula obtained
by adding $\alpha$
as an additional $C^i$ clause into $\varphi$, $i\geq1$. On the other hand, we denote $\varphi[\alpha=0]$ as the formula obtained
by assigning all literals in $\alpha$ to $0$.
\end{definition}

\noindent
The above are common kinds of branching techniques that we will deploy in our paper. 
An example of branching $\alpha=1$ can be seen in this example :
Suppose we have $\alpha \subset C$, for some $C^1$ clause $C$. Then if $\alpha$ is added as new $C^1$ clause into $\varphi$,
we know that the literals in $C-\alpha$ must be assigned to $0$.

\medskip
\noindent\textbf{Resolution.}
Resolution is a technique that allows us to remove variables that appear both
negated and unnegated, at the same time,
preserving the overall satisfiability of the formula. We use it only in the
case that the variable occurs positive and negative in $C^1$ clauses
$C = (\alpha \vee x)$ and $C' = (\beta \vee \neg x)$, where $x$ is th variable, 
and $\alpha$, $\beta$ are subclauses. Now we can replace every clause $C^h$
clause $(x \vee \delta)$ by $(\beta \vee \delta)$ which is also a $C^h$ clause
for the same $h$
and every $C^\ell$ clause $(\neg x \vee \gamma)$ by $(\alpha \vee \gamma)$
which is also a $C^\ell$ clause for the same $\ell$.

\section{G2XSAT (Polynomial Space)}

\noindent
In this section, we give a $O(1.3674^n)$ time and polynomial space algorithm to solve G$2$XSAT.
We will first introduce the design of our nonstandard measure, followed by the debt-taking technique,
then the algorithm and its analysis.

\subsection{Nonstandard measure for the G2XSAT algorithm}

\noindent
We say that a variable $x$ strongly appears in a clause $C$ iff
the truth-value of $C$ depends properly on the value of $x$.
So $x$ strongly appears in the $C^1$ clause $(x \vee y \vee z)$ and 
the $C^2$ clause $(x \vee x \vee \neg x \vee y)$ but not
in the $C^1$ clause $(x \vee \neg x \vee y \vee z)$, as here the truth-value
of $x$ cancels out. Therefore letting $y=0, z=0$ and removing the $C^1$ clause
$(x \vee \neg x \vee y \vee z)$ does not make the measure of $x$ to go up.
Note that whenever in a situation in the algorithm G$2$XSAT
below Lines 1--4 do not apply,
then for each $C^1$ clause $C$ and each variable $x$, $x$ appears in $C$
iff $x$ strongly appears in $C$; hence when handling Lines beyond Line 4,
these two notions are equivalent.

Let $v_i$ be a variable in $\varphi$. We define the weight $w_i$ for $v_i$ as
follows:
\[
 w_i = 
\begin{cases}
    1,& \text{if $v_i$ strongly appears only in $C^2$ clauses of length at least 4;}\\
    0.8039, & \text{otherwise}
\end{cases}
\]
The value $0.8039$ is an optimal value chosen by our linear
search computer program to bring down 
the overall runtime of the algorithm to as low as possible.
In addition, note that our measure $\mu = \sum_{i} w_i \leq n$.
Therefore, this gives us $O(c^\mu) \subseteq O(c^n)$.

\begin{exmp}
Say we have a clause $C^1 = (x \vee y \vee z)$ and a clause 
$C^2 = (x \vee a \vee b \vee c \vee d)$. Then the weights for each variable are as such :
weights for $a=b=c=d=1$, while $x=y=z=0.8039$. Even though the variable $x$ appears in the above
two clauses, as long as it appears in a $C^1$ clause, we assign that lower weight to it.
\end{exmp}

\begin{remark}
In the course of the algorithm, it might occur that a $C^1$-clause is
dissolved and therefore some surviving variables in the clause are increasing
their weight from $0.8039$ to $1$. This can happen when a variable
receives a new value and only two literals remain in the clause,
two variables are linked or there is a resolution. The following
description shows what happens.
\begin{enumerate}
\spaceo
\item Dissolving a clause by setting the value of a variable:
      The first case is that there is a variable $x$ and quite a
      number of clauses of the form $x \vee y_h \vee \neg y_h$
      for variables $y_1,y_2,\ldots,y_k$; here one can conclude
      $x=0$ and simplify accordingly, but then the $k$ $C^1$ clauses
      are removed and this might cause $y_1,y_2,\ldots,y_k$
      no longer to appear in a $C^1$-clause. This problem is
      solved by defining in the measure that only variables which
      appear strongly in a $C^1$ clause are receiving the lower weight
      and the $y_h$ in the above clauses satisfy that $y_h \vee \neg y_h$
      is always $1$ and thus the value $x \vee y_h \vee \neg y_h$
      does not depend on the actual choice of the value of $y_h$,
      so $y_h$ does not appear strongly in the clauses. In the
      case that a literal in a three literal $C^1$-clause is set
      to $1$, all other literals are set to $0$ and thus no gain of
      weight is left. The remaining case is that a literal is set to
      $0$ and only one or two literals remain. The case of one literal
      assigns to that one the value $1$ and again no variable is left
      to gain weight. The case of two literals either leads to linking
      (see next item) or to an elimination of variables as above or
      to the clause being unsatisfiable (like $y \vee y$ for a $C^1$ clause).
\item The second case is linking. Here one concludes that a variable
      $y$ is becoming either $z$ or $\neg z$ and that therefore a clause
      simplifies accordingly. For linked variables, in cases where the
      value of the variable can go up, the verification assigns only a
      change of measure of $2 \cdot 0.8039-1 = 0.6078$ as two variables
      of weight $0.8039$ are replaced by a variable of weight $0.6078$,
      in order to address the fact of weight gain of the resulting variable.
      However, in some cases one can avoid the weight gain and then this
      is stated explicitly in the verification.
\item Resolution is used only in Line 14 of the G$2$XSAT algorithm. Prior to this
      line, Line 9 has removed all overlaps between $C^1$ formulas and
      Lines 10 and 11 have removed all multiple occurrences of variables
      in $C^2$ clauses and Line 12 has removed multiple overlaps between
      $C^1$ and $C^2$ clauses. There are only three subcases where clauses
      of the form $z \vee \beta$, $\neg z \vee \gamma$ are resoluted and
      the literal $z$ is the result of two literals getting the opposite
      value due to a case distinction and linking
      (they had different variables before). As the $C^1$ clauses have
      no overlap before the linking and the surviving part of the $C^2$
      clause, which might be downgraded to a $C^1$-clause, does no longer
      depend on $z$ and thus $z$ does not strongly appear in it,
      the only variables affected by the resolution which might gain weight
      are those in $\beta$ and $\gamma$, however, as none of these variables
      occurs twice in $\beta \vee \gamma$ (neither as $u \vee u$ nor as
      $u \vee \neg u$), these variables also continue to strongly appear
      in $\beta \vee \gamma$ which is a $C^1$ clause; hence the weight of
      these variables does not change. For that reason, the applications of
      resolution {\em inside the algorithm G$2$XSAT} do not increase the
      weight of any variable. For more details, see the verification of the
      branching numbers for Line 14.
\end{enumerate}
\end{remark}

\subsection{Debt Taking}

\noindent
This is a method where one can borrow - up to constant size - measure in order to get a better branching number for a splitting; 
for consistency reasons, it is however required that immediate follow-up operations in those branches where the debts were taken 
out restore the debt to the original value. Note that the debt is taken out by the outgoing branch and does therefore not affect 
the other branches; however, all branches in the follow-up operations must restore the debt. 
In some cases, several branches take out a debt simultaneously which then requires also for all of these branches a 
subsequent action to immediately repay the debt. 

This technique is a state based measure \cite{CKX05,Wa04}.
One can define a state $S_0$ where it is debt-free. For each
branching rules $r$ that borrows up to a certain constant $c_r$, we can define a state $S_r$ 
such that when $S_0$ transitions there, we add $c_r$ to our change of measure.
Likewise, when $S_r$ transitions to $S_0$, we minus $c_r$ from our change of measure. 
The debt taking technique free us from the trouble of keeping track of the different 
states we are in, by just repaying the debt immediately after we borrow. This simplifies our writing. 
For more information on state based measures, one may refer to \cite{Wa07,S08}.

\subsection{Algorithm (Overview)}

\noindent
In this section, we give the algorithm to solve G$2$XSAT. It takes in a formula $\varphi$ in CNF and
outputs a value of $1$ or $0$; where $1$ denotes satisfiable while $0$ otherwise. If there are no heavy variables
in the formula, then we can solve them in polynomial time.
Therefore, the idea here is to remove all heavy variables so that we can solve G$2$XSAT in polynomial time.
The algorithm goes through line by line in this order of decreasing priority. Line 1 has the
highest priority, followed by 2 and so on. When we are at a certain line of the algorithm, we will assume that the 
earlier lines of the algorithm no longer apply. We will call our recursive (branch and bound) algorithm G$2$XSAT(.). 

\medskip
\noindent
Algorithm : G$2$XSAT \\
Input : $\varphi$ \\
Output : $1$ for satisfiable $\varphi$ and $0$ for unsatisfiable $\varphi$.

\begin{enumerate}
\spaceo
\item If there is a clause $C^i$, $i \in \{1,2\}$, that is not satisfiable, then return $0$.
\item If there is a $C^i$ clause $C=(1 \vee \alpha)$ or $C=(x \vee \neg x \vee \alpha)$, then we do the following :
If $i=1$, then $\alpha=0$. Return G$2$XSAT$(\varphi[\alpha=0])$ and drop the clause $C$. Else if $i=2$, then
drop $C$ from $C^2$ to $C^1$ clause, and let $C=\alpha$ and update $\varphi$ as $\varphi'$. Return G$2$XSAT$(\varphi')$.
\item If there is a $C^i$ clause $C=(0 \vee \alpha)$, then update $C=\alpha$ and update $\varphi$ as $\varphi'$.
Return G$2$XSAT$(\varphi')$.
\item If there is a clause $C^i$ containing a literal $x$ appearing $j$ times, $j>i$, then 
return G$2$XSAT$(\varphi[x=0])$. 
\item If there is a $C^2$ clause $C$ with $k$ literals, each appearing exactly twice, $C=(x_1 \vee x_1 \vee x_2 \vee x_2 \vee ... \vee x_k \vee x_k)$, 
then drop it to a $C^1$ clause $C=(x_1 \vee x_2 \vee ... \vee x_k)$, with each of the $k$ literals appearing exactly once. Update
$\varphi$ as $\varphi'$ and then return G$2$XSAT$(\varphi')$.
\item If there is a $C^1$ clause $C$ such that $|C|=2$, then let $x,y$ be the literals in $C$.
		Return G$2$XSAT$(\varphi[x=\neg y])$.
\item If there is a $C^2$ 3-literal clause $C=(x \vee y \vee z)$,
then replace $C$
by the $C^1$ clause $C'=(\neg x \vee \neg y \vee \neg z)$,
where $C'$ is a $C^1$ 3-literal clause. Update $\varphi$ as $\varphi'$ and then return G$2$XSAT$(\varphi')$.
\item If $C$ is a $C^1$ clause with $|C|\geq4$, then we choose $x,y$ in $C$
and branch $x=\neg y$ and $x=y=0$. Return G$2$XSAT($\varphi[x=\neg y]$) $\vee$ G$2$XSAT($\varphi[x=y=0]$). 
\item If $C$ and $C'$ are $C^1$ 3-literal clauses such that $Var(C)\cap Var(C') \neq \emptyset$. Then we either simplify
this case further or we branch $x=1$ and $x=0$. If we simplify this case further,
then let $\varphi'$ be the updated formula after simplifying. Return G$2$XSAT($\varphi'$). If we are branching $x=1$
and $x=0$, then return G$2$XSAT($\varphi[x=1]$) $\vee$ G$2$XSAT($\varphi[x=0]$). 
\item If $C$ is a $C^2$ clause with at least two literals $x$ and $y$ appearing twice. We either simplify some of these
cases further, or we branch $x=\neg y$ and $x=y=0$. If we simplify some of the
cases, then let $\varphi'$ be the new formula after updating $\varphi$ and we return G$2$XSAT($\varphi'$). Else if we branch,
then return G$2$XSAT($\varphi[x=\neg y]$) $\vee$ G$2$XSAT($\varphi[x=y=0]$). 
\item If $C$ is a $C^2$ clause containing exactly one literal $x$ appearing twice. We either branch or simplify
the cases further. If we simplify some of the cases, then let $\varphi'$ be the new formula after 
updating $\varphi$ and we return G$2$XSAT($\varphi'$). Else if we branch, we choose a literal 
$u$ to branch $u=1$ and $u=0$ and then return 
G$2$XSAT($\varphi[u=1]$) $\vee$ G$2$XSAT($\varphi[u=0]$).
\item If there is a $C^1$ clause $C$ and $C^2$ clause $C'$ such that $|Var(C)\cap Var(C')|\geq2$. Then we can update $\varphi'$
from $\varphi$ after simplifying some cases. Return G$2$XSAT($\varphi'$). 
\item If there are $C^2$ 4-literal clause $C=(x \vee y \vee z \vee w)$. We choose variables $u_1$,$u_2$,...$u_k$ and branch
them with values $a_1$, $a_2$, ..., $a_k$, for some $k$. Return G$2$XSAT($\varphi[u_1=a_1]$) $\vee$ 
G$2$XSAT($\varphi[u_2=a_2]$) $\vee ... \vee$ G$2$XSAT($\varphi[u_k=a_k]$). More details on the choosing and branching
of variables in the next section. 
\item If there are $C^1$ clauses $C$ and $C^2$ clauses $C'$ such that $|Var(C)\cap Var(C')|=1$. Then choose a variable $x$
and branch $x=1$ and $x=0$. Return G$2$XSAT($\varphi[x=1]$) $\vee$ G$2$XSAT($\varphi[x=0]$).
\item  If there are $C^2$ clauses $C$ and $C'$ such that $|Var(C) \cap Var(C')|\geq2$, then we can either simplify further
or we have to branch some of these cases. If we are simplifying some of these cases, then let $\varphi'$ be the new
formula from $\varphi$. Return G$2$XSAT($\varphi'$). If we are branching, then choose variables $u_1$,$u_2$,...$u_k$ and branch
them with values $a_1$, $a_2$, ..., $a_k$, for some $k$. Return G$2$XSAT($\varphi[u_1=a_1]$) $\vee$ 
G$2$XSAT($\varphi[u_2=a_2]$) $\vee ... \vee$ G$2$XSAT($\varphi[u_k=a_k]$).
\item In this line, we deal with heavy variable $x$ in the formula that matches the specific subcases (more details later)
and we branch $x=1$ and $x=0$.
Return G$2$XSAT($\varphi[x=1]$) $\vee$ G$2$XSAT($\varphi[x=0]$). 
\item Brute force the remaining heavy variables in the formula. 
\item Solve the rest of the formula in polynomial time. Return $1$ if it is satisfiable, else return $0$.
\end{enumerate}

\noindent
We describe the algorithm here. Line 1 of the algorithm says that if there is any clause that is unsatisfiable, return 0. 
Line 2 of the algorithm deals with constants $1$ appearing in a clause. If it is a $C^1$ clause, then by definition, all
literals in $\alpha$ can be assigned $0$. On the other hand, if it is a $C^2$ clause, then $\alpha$ becomes the new $C^1$ clause.
Line 3 of the algorithm deals with constants $0$ appearing in a clause. We can then remove $0$ from the clause safely. After Line 3,
there will be no more constants in any clause. In Line 4 of the algorithm, if there are literals $x$ occuring $j$ times in a $C^i$ clause,
where $j>i$, then set $x=0$. This is because we are not allowed to over-satisfy by definition of the problem. So after this, every literal
can only appear at most $i$ times, in a $C^i$ clause. In Line 5 of the algorithm, we deal with $C^2$ clauses with every literal appearing twice.
This is equivalent to solving $C^1$ clauses with each literal appearing exactly once in it. In Line 6 of the algorithm, if there a 2-literal $C^1$
clause containing the literals $x$ and $y$, then we can just set $x=\neg y$. After Line 6, all $C^1$ clauses must be at least $k$-literal, 
where $k\geq3$. In Line 7 of the algorithm, when we have a 3-literal $C^2$ clause say $C=(x \vee y \vee z)$, then we can set
$C=(\neg x \vee \neg y \vee \neg z)$ to be a $C^1$ clause. This is valid since $C$ was originally was a $C^2$ clause, requiring exactly
2 literals to be assigned $1$ and a literal assigned $0$. Now if we were to negate all the literals, then now exactly a literal needs to
be assigned $1$ and 2 literals assigned $0$. After this, all $C^2$ clause must be at least a $k$-literal clause, where $k\geq4$. Note
that Lines 1 to 7 of the algorithm are simplification rules. These will only take polynomial time. 

From Line 8 of the algorithm onwards, these lines contribute to the exponential growth of the algorithm. Hence, an indepth analysis
of their time complexity will be studied in greater detail in the next section. In Line 8 of the algorithm, we deal with $C^1$ $k$-literal
clauses, where $k\geq4$. This means that after this line, only $C^1$ 3-literal clauses are left in the formula with $C^2$ clauses.
We delay handling of $C^1$ 3-literal clauses because directly handling them now would increase the bottleneck of the entire algorithm. 
In Line 9 of the algorithm, we deal with two $C^1$ 3-literal clauses that have variables in common. After this, all $C^1$ 3-literal clauses $C$
can only appear with $C^2$ clauses $C'$; $Var(C) \cap Var(C') \neq \emptyset$. In Line 10 of the algorithm, we deal with
$C^2$ clauses containing at least two literals appearing twice. In Line 11 of the algorithm, we deal with $C^2$ clauses that contains a literal
appearing twice in the clause. After Line 11, any literal appearing in any $C^2$ clause must only appear once. Line 12 of the algorithm
deals with $C^1$ clause $C$ and $C^2$ clause $C'$ such that $|Var(C) \cap Var(C')|\geq2$. After Line 12 of the algorithm, we know that
if any $C^1$ clause $C$ and $C^2$ clause $C'$ have variables in common, then we have $|Var(C) \cap Var(C')|\leq1$. In Line 13 of the 
algorithm, we deal with $C^2$ 4-literal clause. After which, all $C^2$ clauses must be $k$-literal clauses, where $k\geq5$. In Line 14
of the algorithm, we deal with $C^1$ clause $C$ and $C^2$ clause $C'$ such that $|Var(C) \cap Var(C')|=1$. After which, we will no 
longer deal with $C^1$ 3-literal clauses. In Line 15 of the algorithm, we deal with $C^2$ clauses $C$ and $C'$ such that
$|Var(C) \cap Var(C')|\geq2$. After which, we know that given any two $C^2$ clauses $C$ and $C'$, we must have that
$|Var(C) \cap Var(C')|\leq1$. We are now ready to deal with heavy variables in the formula. Line 16 and 17 of the algorithm deals
with heavy variables in the formula. We branch certain cases in Line 16, and brute force the remaining heavy variables in Line 17.
After which, we can solve the remaining formula in polynomial time and then return $1$ if satisfiable and $0$ otherwise.

Note that we have covered every case above and therefore shows the correctness of our branch and bound algorithm. 

\subsection{Analysis of Algorithm}

Note that whenever we perform linking in a $C^1$ clause, the surviving variable may have its weight increased 
from 0.8039 to 1 in the worst case scenario. This is due to the fact that the variable might not appear in any other $C^1$ clauses, hence
we need to account for this increase in weight. For example, if we have $(x \vee y \vee z)$ and when $x=0$,
we have that $y=\neg z$ and we remove $y$. Now the surviving variable $z$
may not appear in any more $C^1$ clauses and its weight may increase from $0.8039$ to $1$, a decrease
in measure of $0.1961$. On the other hand, we remove $y$, which has weight $0.8039$. Therefore, the
net change of measure is $0.8039-0.1961=0.6078$ whenever we perform linking in a $C^1$ clause.

In this section, we will present the analysis of the algorithm from Line 8 of the algorithm onwards. Lines 1 to 7
are simplification rules which only take polynomial time. Lines 8 onwards are branching rules which contribute to the 
exponential growth of the algorithm. Finally, Line 18 also takes polynomial time once no heavy variables exist in the
formula. Therefore, we will analyse Lines 8 to 17 of our algorithm here. Note that from Line 4 onwards, the notion
of appear and strongly appear coincide.

\begin{enumerate}
\spaceo
\setcounter{enumi}{7}
\item If $C$ is a $C^1$ clause with $|C|\geq4$, then we choose $x,y$ in $C$
and branch $x=\neg y$ and $x=y=0$.

We consider the branching factor for each $|C|\geq4$. 
\begin{itemize}
\spaceo
\item If $|C|=4$. Then let $x,y,z,w$ be the 4 literals in $C$. We branch $x=\neg y$ and $x=y=0$.
When $x=\neg y$, then $z=w=0$ and thus we remove $z,w$. 
This gives us a change of measure of $2\times0.8039+0.6078=2.2156$.
When $x=y=0$, we link $z=\neg w$. This gives us a change of measure of $2\times0.8039+0.6078=2.2156$.
The branching factor is $\tau(2.2156,2.2156)=1.3674$.
\item If $|C|\geq5$. Choose two literals $x,y$ and branch $x=\neg y$
and $x=y=0$.
If $x=\neg y$, then all the other literals in $C$ must be $0$,
removing at least 3 variables here.
This gives us a change of measure of $3\times0.8039+0.6078=3.0195$.
On the other hand, when $x=y=0$,
we remove only 2 variables $x,y$, giving us a change of measure of
$2\times0.8039=1.6078$. Therefore,
the branching factor for this case is $\tau(3.0195,1.6078)=1.3633$.
\end{itemize}
After this line of the algorithm has been executed, the only $C^1$ clauses left are 3-literal clauses. We have
factored in the increase of weights due to linking variables while branching in our branching factors above.

\item If $C$ and $C'$ are $C^1$ 3-literal clauses such that $(Var(C)\cap Var(C')) \neq \emptyset$. Then
do the following:
\begin{itemize}
\item $|Var(C) \cap Var(C')|=1$. If we have $C=(x \vee \alpha)$ and
$C'=(\neg x \vee \beta)$. Then we branch $x=1$ and $x=0$. When $x=1$,
we remove $x$, $\alpha$ and link two of the variables in $\beta$. This gives
us a change of measure of $3\times0.8039+0.6078=3.0195$.
The case for $x=0$ is symmetric. This gives us a branching
factor of $\tau(3.0195,3.0195)=1.2581$.
Else, we must have $C=(x \vee \delta)$ and $C'=(x \vee \alpha)$.
If $x=1$, then we remove $x$, $\alpha$ and $\delta$.
This gives us a change of measure of 
$5\times0.8039=4.0195$.
If $x=0$, we remove $x$ and link the two variables in $\alpha$ and $\beta$.
This gives us $0.8039+2\cdot 0.6078=2.0195$. This gives us a branching
factor of $\tau(4.0195,2.0195)=1.2699$.
\item $|Var(C) \cap Var(C')|=2$. \\
If $C=(x \vee y \vee z)$ and $C'=(x \vee y \vee w)$. Then link $z=w$.
If $C=(x \vee y \vee z)$ and $C'=(x \vee \neg y \vee w)$, then $x=0$.
If $C=(x \vee y \vee z)$ and $C'=(\neg x \vee \neg y \vee w)$,
then $x=\neg y$ and $z=w=0$. 
\item $|Var(C) \cap Var(C')|=3$. The case $C=(x \vee y \vee z)$ and
$C'=(\neg x \vee \neg y \vee \neg z)$, 
$C=(\neg x \vee y \vee \neg z)$ and $C'=(x \vee \neg y \vee z)$,
$C=(x \vee y \vee z)$ and $C'=(\neg x  \vee y \vee z)$ are not satisfiable.
If we have $C=(x \vee y \vee z)$ and $C'=(x \vee \neg y \vee \neg z)$,
then $y=\neg z$ and $x=0$.  
\end{itemize} 
After this line of the algorithm, $C^1$ 3-literal clauses can only appear with
$C^2$ clauses. Here, the increase in weights due to linking have also been factored in
our branching factors. 

\item If $C$ is a $C^2$ clause with at least two literals $x$ and $y$ appearing twice. We simplify some of these
cases further. 

Line 5 of the algorithm helps us to handle the following cases: $C=(x \vee x \vee y \vee y)$, 
$C=(x \vee x \vee y \vee y \vee z \vee z)$ and $C=(x \vee x \vee y \vee y \vee z \vee z \vee w \vee w)$. Therefore,
it suffices to only consider other cases. Note that $|C|\geq5$.
\begin{itemize}
\item $|C|=5$. $C=(x \vee x \vee y \vee y \vee z)$. Then $z=0$. 
\item $|C|=6$. $C=(x \vee x \vee y \vee y \vee z \vee w)$. Then we must have that $w=z$. 
\item $|C|=7$ and $C=(x \vee x \vee y \vee y \vee z \vee z \vee w)$. Then $w=0$. 
\item In all other subcases we have at least five variables.
      We branch $x = \neg y$ and $x=y=0$.
      As the weight of each variable is at least $0.8039$ (it is possible that they
	may appear in $C^1$ clauses at this point in time) and as the measure
      of $x,y$ might go up when they get linked,
      we apply the same analysis as in the case $|C| \geq 5$ of Line 8:
      If $x=\neg y$, then all the other literals in $C$ must be $0$,
      removing at least 3 variables here.
      This gives us a change of measure of $3\times0.8039+0.6078=3.0195$.
      On the other hand, when $x=y=0$,
      we remove only 2 variables $x,y$, giving us a change of measure of
      $2\times0.8039=1.6078$. Therefore,
      the branching factor for this case is $\tau(3.0195,1.6078)=1.3633$.
\end{itemize}

After this line of the algorithm, every $C^2$ clause must contain at most a literal appearing twice in it. 
The increase in weights due to linking variables have been factored in our branching factors. 

\item If $C$ is a $C^2$ clause containing exactly one literal $x$ appearing twice. We either branch or simplify
the cases further.

Note that by the previous line of the algorithm, every $C^2$ clause can only have at most one literal 
appearing twice in it.
\begin{itemize}
\item $|C|=3$. Then $C=(x \vee x \vee y)$. Then $x=1,y=0$. 
\item $|C|=4$. Then $C=(x \vee x \vee y \vee z)$. Then $y=z$. 
\item $|C|=5$. Then $C=(x \vee x \vee y \vee z \vee w)$. We deal with this case by case.
\begin{itemize}
\item
Now if all the variables in $C$ have weight 1. We branch $x=1$ and $x=0$. If $x=1$, we remove
all 4 variables, giving us a change of measure of 4. If $x=0$, then we remove $x$, and the other
variables drop their weight from $1$ to $0.8039$, giving us a change of measure of 
$1+3\times(1-0.8039)=1.5883$. This gives us $\tau(4,1.5883)=1.3051$.
\item
For the next few cases, we must have that some variables in $C$ have weight 0.8039. This means
that some variables in $C$ appear in a $C^1$ 3-literal clause. Let $C'$ be the $C^1$ 3-literal clause.
We enumerate the cases here.
\item
$C'=(\neg x \vee u \vee v)$. In this case, $u,v$ can be any literal (including $y,z,w$). Then we branch
$x=1$ and $x=0$. We treat all the variables here as having weight $0.8039$ to upper bound our
branching factor. When $x=1$, we remove all 4 variables in $C$, giving us a change of measure 
of $4\times0.8039=3.2156$. When $x=0$, we remove the literals $x,u,v$, giving us a change
of measure of $3\times0.8039=2.4117$. This gives us a branching factor of 
$\tau(3.2156,2.4117)=1.2817$. For the remaining cases below, we will assume that 
$\neg x$ does not occur in $C'$. \\
Now, we deal with the case that $|Var(C)\cap Var(C')|\geq2$. Now the literal $u$ can be any literal for 
the cases below. \\
$C'=(\neg y \vee \neg z \vee u)$. Then $x=1$ cannot happen,
else some clauses are not satisfiable. Therefore $x=0$. \\
$C'=(\neg y \vee z \vee u)$. If $u$ and $w$ are
the same variable then drop $C$ and add in their sum which is
the $C^2$ clause $(x \vee x \vee z \vee z \vee w \vee w)$, 
which may then be further simplified to $C^1$ clause $(x \vee z \vee w)$;
one can afterwards conclude that $\neg y = x$ and simplify further.
If $u$ and $w$ are different variables, one adds again $C$ and $C'$
and gets the $C^2$ clause $(x \vee x \vee z \vee z \vee u \vee w)$ which
allows to link $u = w$ and thus leads to some simplification. \\
$C'=(y \vee z \vee u)$. Now, if $(y \vee z) =1$, then $x=0$ and $w=1$.
Else if $y=z=0$, then $x=1$ and $w=0$. Therefore, in either case, we see that
$x =\neg w$. \\
$C'=(x \vee \neg y \vee u)$. Then we know that if $x=1$, then $y=0$ and $\neg y=1$, 
which makes the $C'$ not satisfiable. Therefore, $x=0$. \\
$C'=(x \vee y \vee z)$. Then $C' \subset C$ and therefore, $(x \vee w)=1$. 
Therefore, $x=\neg w$.
$C'=(x \vee y \vee u)$. The literal $u$ is not in $\{x,y,z,w,\neg x, \neg y,\neg z,\neg w\}$.
Then we branch $u=1$ and $u=0$.
If $u=1$, then $x=y=0$, and $z=w=1$. This gives 
us a change of measure of $5\times0.8039=4.0195$.
If $u=0$, we remove $u$ and then
$x=\neg y$. This gives us a change of measure of $0.8039+0.6078=1.4117$.
This gives us a 
branching factor of $\tau(4.0195,1.4117)=1.3220$.
\item
For the remaining cases, we deal with $|Var(C')\cap Var(C)|=1$. \\
$C'=(x \vee u \vee v)$. Note that the literals $u,v \notin \{x,y,z,w,\neg x,\neg y,\neg z,\neg w \}$.
We branch $x=1$ and $x=0$. Here the worst case is that $y,z,w$ 
do not appear in $C^1$ 3-literal clauses, else when $x=1$, we can factor in additional measure via linking.
Therefore, when $x=1$, $u,v,y,z,w=0$, giving us a change of measure of $6\times0.8039=4.8234$.
When $x=0$, then $u=\neg v$, giving us a change of measure of $0.8039+0.6078=1.4117$. This gives a
branching factor of $\tau(4.8234,1.4117)=1.2852$. \\
Finally, we have the case that $x$ does not appear in any other $C^1$ 3-literal clauses, but the literals $y,z,w$ may. 
Note that at this point, $|Var(C')\cap Var(C)|=1$, for any 3-literal clause $C'$. Let $k$ be the number of 3-literal
clauses that are neighbours to $C$, each containing one of the variable in $Var(\{y,z,w\})$. 
When $x=1$, then we remove all variables $y,z,w$. Now if there are $k$ many 3-literal clauses, then our change
of measure is $1+k\times(0.8039+0.6078) + (3-k)$. When $x=0$, the change of measure for the variables
not appearing in 3-literal clauses, giving us a change of measure of $1+(3-k)(1-0.8039)$. The worst
case happens when $k=3$, giving us a branching factor of $\tau(5.2351,1)=1.3143$. 
\end{itemize}
\item $|C| \geq 6$ and thus $|Var(C)|\geq 5$. So we analyse the most critical
case $C=(x \vee x \vee y \vee z \vee v \vee w)$. We branch $x=1$ and $x=0$,
but for this case, we need a case-distinction in the analysis of the
situation. \\
If $x$ appears in a $C^1$ clause $C'$, then $x$ must have weight $0.8039$.
We treat all the other
variables in $C$ having weight $0.8039$ to act as an upper bound on our branching factor. 
When $x=1$, we remove at least 5 variables in $C$, giving us a change of measure of $5\times0.8039=4.0195$.
When $x=0$, we remove $x$ and link the 2 variables in $C'$, giving us $0.8039+0.6078=1.4117$. This gives us a
branching factor of $\tau(4.0195,1.4117)=1.3220$. \\
If $x$ does not appear in any $C^1$, then it must have
weight $1$. Now when $x=1$, our change of measure
is $1+4\times0.8039=4.2156$. When $x=0$,
we remove only $x$, giving us a change of measure of $1$.
Therefore, this gives us a branching
factor of $\tau(4.2156,1)=1.3665$.
\end{itemize}

The increase in weights due to linking (for the respective branching cases)
has already been factored in our branching factors.

\item If there is a $C^1$ clause $C$ and a $C^2$ clause $C'$ such that
$|Var(C)\cap Var(C')|\geq2$.

Note that at this point onwards,
all literals appearing in any $C^2$ clause can only appear once
and any $C^1$ clause must be a 3-literal clause at this point.
Note that some of the cases here overlaps with those shown in the previous line,
when dealing with the case $(x \vee x \vee y \vee z \vee w)$. 

$|Var(C)\cap Var(C')|=3$. Let $C'=(x \vee y \vee z \vee \alpha)$. 
\begin{itemize}
\item $C=(x \vee y \vee z)$. Then we can drop $C'$ and add in
the $C^1$ clause $\alpha$.
\item $C=(x \vee y \vee \neg z)$.
Then from these two clauses $C$ and $C'$, we arrive at the
$C^2$ clause $(x \vee x \vee y \vee y \vee \alpha)$ (after removing
redundant terms on both sides). So we replace $C'$ by $(x \vee x \vee y \vee y \vee \alpha)$
and handle the resulting set of clauses as in Line 10 of the algorithm.
\item $C=(x \vee \neg y \vee \neg z)$.
If $x=1$ then $\neg y \vee \neg z = 0$ and $y \vee z=2$, giving that
$C'$ has three literals satisfied, a contradiction. Thus $x=0$
and we can simplify accordingly.
\item $C=(\neg x \vee \neg y \vee \neg z)$.
Then all the literals in $\alpha$ must be assigned $0$.
\end{itemize}

$|Var(C) \cap Var(C')|=2$. Let $C'=(x \vee y \vee \alpha)$.
\begin{itemize}
\item $C=(x \vee y \vee z)$. Therefore, $z=\alpha$ and then 
$(z \vee \neg z)=(\alpha \vee \neg z)$ and $1=(\neg z \vee \alpha)$. This is a $C^1$ clause
which can be added in and we can drop the clause $C'$ to decrease the overall measure. 
\item $C=(x \vee \neg y \vee z)$. Then we can replace $C'$ as $(x \vee x \vee z \vee \alpha)$
and this case will be handled by the earlier line of the algorithm (same literal appearing twice).
\item $C=(\neg x \vee \neg y \vee z)$. Then we can replace $C'$ by the $C^1$ clause $(z \vee \alpha)$,
which will be handled by earlier lines of the algorithm.   
\end{itemize}
After this line, for any $C^1$ 3-literal clause $C$ and $C^2$ clause $C'$, we must therefore have
$|Var(C) \cap Var(C')|\leq1$. In this line, we mainly simplify most of the cases. Hence, no increase
in weights for any of the variables in any of the cases here.

\item If there are $C^2$ 4-literal clause $C=(x \vee y \vee z \vee w)$. We choose specific variable(s)
and branch based on the different case scenario below.

By the previous line of the algorithm, if any $C^1$ clause $C$ and $C^2$ clause $C'$ overlap, then
we must have $|Var(C) \cap Var(C')|=1$. 
\begin{itemize}
\item All variables in $C$ have weight 1. Then we branch $x=\neg y$,$z=\neg w$ and $x=y$,$z=w$,
$y=\neg w$. For the 1st branch, we remove 2 variables of weight $1$, giving us a change of measure of 2.
For the 2nd branch, we remove 3 variables of weight $1$, giving us a change of measure of 3.
This gives us a branching factor of $\tau(2,3)=1.3248$.

\item Exactly 1 of the variable appears in a further 3-literal clause.
We do exactly the same as in the previous case, but make sure that only the
variables of weight 1 get eliminated (replaced by other variables via linking).
Thus again the branching factor is $\tau(2,3)=1.3248$.

\item At least 2 of the variables appears in a further 3-literal clause. Let these two variables be $x,y$
and the two $C^1$ clauses be $(x \vee \alpha)$ and $(y \vee \beta)$, for some subclauses $\alpha$
and $\beta$. We treat all variables here as having weight $0.8039$ to upper bound our
branching factor. Then we branch $x=y=1$, $x=\neg y$ and $x=y=0$. Now, when $x=y=1$, then
all literals in $\alpha$ and $\beta$, $z,w$ are all assigned 0. This gives us a change of measure
of $8\times0.8039=6.4312$. When $x=\neg y$, then $z=\neg w$, removing two variables of weight $0.8039$,
giving us a change in measure of $2\times0.8039=1.6078$. Finally, when $x=y=0$, then $z=w=1$, and 
we can link up the two variables in $\alpha$ and $\beta$, giving us a change of measure of
$4\times0.8039+2\cdot 0.6078=4.4312$. This gives us a branching factor of 
$\tau(6.4312,1.6078,4.4312)=1.3620$. Note that if the above
$C^1$-clauses are $(\neg x \vee \alpha)$ and $(\neg y \vee \beta)$, then one
has a symmetric situation and again the branching factor $1.3620$.
If the $C^1$-clauses are $(\neg x \vee \alpha)$ and $(x \vee \beta)$,
then one has a mixed situation where for $x=1,y=1$ as well as $x=0,y=0$
one eliminates six variables and links two and for the middle case,
one eliminates two variables by linking. This gives
$\tau(6 \times 0.8039+0.6078,6 \times 0.8039+0.6078,1.6078) = 1.3540$.
\end{itemize}
Likewise, the increase of weights due to linking variables have also been factored in the 
respective branching factors above.  

\item If there are $C^1$ clauses $C$ and $C^2$ clauses $C'$ such that $|Var(C)\cap Var(C')|=1$.

By the previous line of the algorithm, all $C^2$ clauses $C'$ must be $|C'|\geq5$. We consider
the different cases below: 
\begin{itemize}
\item There are at least three $C^1$ 3-literal clauses that are neighbours to $C'$.
Let $C'=(x \vee y \vee z \vee \delta)$, for some subclause $\delta$, 
and the other clauses be $(x' \vee \alpha)$, $(y' \vee \beta)$ and $(z' \vee \gamma)$,
where $x' \in \{x,\neg x\}$, $y' \in \{y,\neg y\}$ and $z' \in \{z,\neg z\}$. We choose
two of the literals in $\{x',y',z'\}$ such that they appear as both positive literals or
negative literals. Without any loss of generality, we let the literals be $y$ and $z$ and
we have $(y \vee \beta)$ and $(z \vee \gamma)$ (or alternatively 
$(\neg y \vee \beta)$ and $(\neg z \vee \gamma)$). Then we branch $x=1$ and $x=0$. \\
Case $(x \vee \alpha)$. When $x=1$, all literals in $\alpha$ are assigned $0$.
This gives us a 
change of measure of $3\times0.8039=2.4117$. On the other hand, when $x=0$, we link
up the literals in $\alpha$ to get a change of measure of $1.4117$. For the 1st branch,
we take a debt of $0.98$ to get a branching factor of $\tau(2.4117+0.98,1.4117)=1.3602$. \\
Case $(\neg x \vee \alpha)$. This case is symmetric to the above case. We take a debt 
of $0.98$ when $x=1$. This gives us a branching factor of 
$\tau(1.4117+0.98,2.4117)=1.3346$. \\
Follow up: Note that in both cases above, we took a debt of $0.98$ when $x=1$. We need
to clear this debt in their subsequent child nodes where the debt took place. \\
We branch $y=\neg z$ and $y=z=0$. \\
Case $(y \vee \beta)$ and $(z \vee \gamma)$. \\ 
$|\delta|\geq3$. When $y=\neg z$, then all literals in $\delta$ are assigned 0. 
Then we have the clauses $(\neg z \vee \beta)$ and $(z \vee \gamma)$ where we can apply resolution
to remove $z$. This gives us a change of measure of $5\times0.8039=4.0195$. When $y=z=0$,
we can link the two literals in $\beta$ and $\gamma$, giving us a change of measure of 
$2\times0.8039+2\cdot 0.6078=2.8234$.
The branching factor is $\tau(4.0195-0.98,2.8234-0.98)=1.3365$. \\
$|\delta|=2$. When $y=\neg z$, then the two literals in $\delta$ are assigned 0. 
Then we have the clauses $(\neg z \vee \beta)$ and $(z \vee \gamma)$ where we can apply resolution
to remove $z$. This gives us a change of measure of $4\times0.8039=3.2156$. When $y=z=0$,
we can link the two literals in $\beta$, $\gamma$ and $\delta$, giving us a change of measure of 
$2\times0.8039+3\cdot 0.6078=3.4312$. The branching factor is $\tau(3.2156-0.98,3.4312-0.98)=1.3445$. \\
Case $(\neg y \vee \beta)$ and $(\neg z \vee \gamma)$. \\
When $y=\neg z$, then all literals in $\delta$ are assigned 0, then we remove at least 3 variables.  
Then we have the clauses $(\neg z \vee \beta)$ and $(z \vee \gamma)$ where we can apply resolution
to remove $z$. This gives us a change of measure of $4\times0.8039=3.2156$. On the other hand,
when $y=z=0$, we remove all the literals in $\beta$ and $\gamma$, giving us a change of measure
of $6\times0.8039=4.8234$. This gives us a branching factor of 
$\tau(3.2156-0.98,4.8234-0.98)=1.2635$.
\item There are two $C^1$ 3-literal clauses that are neighbours to $C'$. Suppose we have 
$C'=(x \vee y \vee \delta)$, $|\delta|\geq3$ and $C^1$ clauses $(x \vee \alpha)$, $(y \vee \beta)$. We
branch $x=1$ and $x=0$. When $x=1$, we remove $x$ and all literals in $\alpha$, and all 
variables in $\delta$ drop in measure to $0.8039$ from $1$. This gives us a change of measure of
$3\times0.8039+3\times(1-0.8039)=3$. When $x=0$, we remove $x$ and link the variables in $\alpha$,
giving us a change of measure of $0.8039+0.6078=1.4117$. We take a debt of $0.39$ for the first branch.
This gives us a branching factor of $\tau(3+0.39,1.4117)=1.3604$.  Symmetrically, if we have
$(\neg x \vee \alpha)$. Then branching $x=1$ will give us $0.8039+0.6078+3\times(1-0.8039)=2$,
and branching $x=0$ will give us $3\times0.8039=2.4117$. We again, take a debt of $0.39$ for the
$x=1$ branch. This gives us a branching factor of $\tau(2+0.39,2.4117)=1.3348$. \\
Follow up: In the $x=1$ branch, we are left with a $C^1$ clause $(y \vee \delta)$, with 
$y$ also appearing in another 3-literal clause $(y \vee \beta)$. Now we branch $y=1$ and $y=0$. 
When $y=1$, we remove all the literals in $\delta$ and $\beta$, giving us a change of measure of $6\times0.8039=4.8234$.
When $y=0$, we remove $y$ and link the literals in $\beta$, giving us a change of measure of 
$0.8039+0.6078=1.4117$. Paying off the earlier debt of $0.39$ will give us 
$\tau(4.8234-0.39,1.4117-0.39)=1.3502$. Symmetrically, if we have $(\neg y \vee \beta)$, then 
$y=1$ will give us a change of measure of $4\times0.8309+0.6078=3.8234$. On the other hand,
when $y=0$, this gives us $2.4117$. This gives us $\tau(3.8234-0.39,2.4117-0.39)=1.2973$.
\item There is exactly one $C^1$ 3-literal clause that is a neighbour to $C'$. Suppose we have 
$C'=(x \vee \delta)$, $|\delta|\geq4$, and a $C^1$ clause $(x' \vee \alpha)$, where $x' \in \{x,\neg x\}$. 
We analyse it by different case. \\
Case $|\delta|\geq5$. Suppose we have $(x \vee \alpha)$ and when $x=1$, we remove all literals in 
$\alpha$, and the variables in $\delta$ drop from $1$ to $0.8039$, giving us a change of measure of 
$3\times0.8039+5\times(1-0.8039)=3.3922$. When $x=0$, we remove $x$ and link the literals
in $\alpha$, giving us a change of measure of $0.8039+0.6078=1.4117$. This gives us a branching factor of
$\tau(3.3922,1.4117)=1.3602$. \\
Case $|\delta|=4$. Therefore, all the variables appearing in $\delta$ have weight 1 and we also have
a $C^1$ clause $(x' \vee \alpha)$, where $x' \in \{x,\neg x\}$. If we have $(\neg x \vee \alpha)$,
when $x=1$, we remove $x$, link the literals in $\alpha$, and all the variables in $\delta$ drop
their weight from $1$ to $0.8039$, giving us a change in measure of $0.8039+0.6078+4\times(1-0.8039)=2.1961$.
When $x=0$, all literals in $\alpha$ are assigned 0, giving us a change in measure of $3\times0.8039=2.4117$.
This gives us a branching factor of $\tau(2.1961,2.4117)=1.3514$. If we have $(x \vee \alpha)$, 
if we have $x=1$, all the literals in $\alpha$ are assigned 1, and the variables in $\delta$ will drop in measure,
giving us $3\times0.8039+4\times(1-0.8039)=3.1961$. When $x=0$, we remove $x$ and link up the 
literals in $\alpha$, giving us a change of measure of $0.8039+0.6078=1.4117$. 
For the $x=0$ branch, we take a debt of $0.2$ to give us the branching factor of
$\tau(3.1961,1.4117+0.2)=1.3499$. \\
Follow up: For the $x=0$ branch, we are left with a $C^2$ clause $\delta$ with all variables having weight 1.
From Line 13, we will have a branching factor of $\tau(3-0.2,2-0.2)=1.3587$. 
\end{itemize}
Note that for the cases above that uses linking, the increase in weights have been factored in the branching factors above.
For the cases where we first use linking, and then followed by resolution, no increase in weights for the variables after
resolution due to the fact that we only apply it on $C^1$ 3-literal clauses.
Resolution will increase the length of the clause,
but not the weights in the clause due to how we define our nonstandard measure. 

\item If there are $C^2$ clauses $C$ and $C'$
such that $|Var(C) \cap Var(C')|\geq2$, then we do the following:

From this point onwards, all variables appearing in $C^2$ must all have weight 1. From Line 13, we know
that both $|C|\geq5$ and $|C'|\geq5$. All clauses are $C^2$ clauses by default unless specifically mentioned.
\begin{itemize}
\item We deal with some special cases here. \\
$C \subset C'$, then set all literals in $C'-C$ to be 0. \\
$C=(x \vee \alpha)$ and $C'=(\neg x \vee \alpha \vee \beta)$.
Then $\alpha$ combines with both $x, \neg x$ in $C^2$ clauses
and is just either $0$ or $1$; thus we can conclude that
$x=1$ to get $C$ satisfied and simplify according. \\
$C=(x \vee y \vee \alpha)$ and $C'=(\neg x \vee \neg y \vee \alpha \vee \beta)$.
We can derive $x \vee y = \neg x \vee \neg y \vee \beta$ and
by adding $\neg x \vee \neg y$ on both sides will give us
$(\neg x \vee \neg x \vee \neg y \vee \neg y \vee \beta)=2$;
that is, a $C^2$ clause
$(\neg x \vee \neg x \vee \neg y \vee \neg y \vee \beta)$.
This $C^2$ clause replaces $C'$ in $\varphi$ and it will
be handled as in Line 10 of the algorithm. \\
$C=(x \vee y \vee z \vee \alpha)$ and
$C'=(\neg x \vee \neg y \vee \neg z \vee \alpha \vee \beta)$.
The sum of these two gives the $C^1$ clause $(\alpha \vee \alpha \vee \beta)$
and we set all literals in $\alpha$ to $0$ and replace $C'$ by
the $C^1$ clause $\beta$.
\item If we have $|Var(C)-Var(C')|=1$ or $|Var(C')-Var(C)|=1$,
then we deal it case by case. \\
$C=(x \vee \alpha)$ and $C'=(\alpha \vee \beta)$.
One can derive that $x=\beta$. Therefore,
adding $\neg x$ to both sides, we have $1=(\neg x \vee \beta)$.
Therefore, $(\neg x \vee \beta)$ is
a $C^1$ clause. We can add this in to reduce the measure
of the overall formula. \\
$C=(x \vee y \vee \alpha)$ and $C'=(\neg y \vee \alpha \vee \beta)$. Note that $|\alpha|\geq3$.
We branch $\alpha=1$ and $\alpha=0$. When $\alpha=1$, then $x=\neg y$,
removing 1 variable here.
When $\alpha=0$, then $x=y=1$, which means we remove at least 5 variables here. This gives us 
$\tau(1,5)=1.3248$. \\
$C=(x \vee y \vee z \vee \alpha)$ and
$C'=(\neg y \vee \neg z \vee \alpha \vee \beta)$. Note that $|\alpha|\geq2$.
Let $\gamma = \alpha \vee x$. Note that $\gamma=2$ is impossible,
as otherwise $y=z=0$ and therefore
all literals in $\alpha$ and $\beta$ must be assigned 0, a contradiction.
Therefore, we can only branch $\gamma=1$ and $\gamma=0$.
When $\gamma=1$, then $y=\neg z$, removing 1 variable here.
When $\gamma=0$, then all literals in $\gamma$ are assigned $0$,
allowing us to remove at least 3 variables here.
In addition, $y=z=1$. Therefore, we remove 5 variables here,
giving us $\tau(1,5)=1.3248$. \\
$C=(x \vee y \vee z \vee w \vee \alpha)$ and
$C'=(\neg y \vee \neg z \vee \neg w \vee \alpha \vee \beta)$. Then
$\alpha=0$; as $|\alpha| \geq 1$ by the fact that $|C| \geq 5$, so no branching is required here. \\
$C=(x \vee y \vee z \vee w \vee v \vee \alpha)$ and $C'=(\neg y \vee \neg z \vee \neg w \vee \neg v \vee \alpha \vee \beta)$.
Then $\alpha=\beta=x=0$.
\item $|Var(C) \cap Var(C')|=2$. We can either have $C=(x \vee y \vee \alpha)$
and $C'=(x \vee y \vee \beta)$ (first case),
$C=(x \vee y \vee \alpha)$ and $C'=(x \vee \neg y \vee \beta)$ (second case), 
$C=(x \vee y \vee \alpha)$ and $C'=(\neg x \vee \neg y \vee \beta)$ (third case). Note that $|\alpha|\geq3$ and $|\beta|\geq3$.
For the first case, we branch $x=y=1$ (first branch), $x=\neg y$ (second branch) and $x=y=0$ (third branch). \\
The first branch removes all variables, second branch removes $x$, and drops the rest of the variables in $\alpha$
and $\beta$ from $1$ to $0.8039$ and finally the third branch removes $x,y$ and we are left with $C^2$
clauses $\alpha$ and $\beta$.

Case $C=(x \vee y \vee \alpha)$ and $C'=(x \vee y \vee \beta)$. \\
$|\alpha|=|\beta|=3$. First branch removes all 8 variables. Second branch gives us a change of measure of 
$1+6\times(1-0.8039)=2.1766$. For the third branch, $\alpha$ and $\beta$ 
can be downgraded to a $C^1$ 3-literal clause by Line 7 of the algorithm, giving us a change of measure of 
$2+1.1766=3.1766$. This gives us $\tau(8,2.1766,3.1766)=1.3485$. \\
$|\alpha|=3,|\beta|=4$. First branch removes all 9 variables. Second branch gives us $1+7\times(1-0.8039)=2.3727$.
Third branch gives $2+3\times(1-0.8039)=2.5883$. This gives $\tau(9,2.3727,2.5883)=1.3582$. \\
$|\alpha|=|\beta|=4$. First branch removes 10 variables. Second branch gives us $1+8\times(1-0.8039)=2.5688$.
Third branch gives us $2$. We take a debt of $0.4$ for the third branch, this gives us $\tau(10,2.5688,2.4)=1.3496$.
The debt of $0.4$ will repaid by the clauses $\alpha$ and $\beta$, each repaying $0.2$. From Line 13, this gives 
$\tau(3-0.2,2-0.2)=1.3586$. \\
$|\alpha|\geq4,|\beta|\geq5$. First branch removes 11 variables. 
Second branch gives $1+9\times(1-0.8039)=2.7649$. Third branch gives $2$. This gives $\tau(11,2.7649,2)=1.3611$.

Case $C=(x \vee y \vee \alpha)$ and $C'=(\neg x \vee \neg y \vee \beta)$. \\
With the earlier cases not applying, we know that $|Var(\alpha) - Var(\beta)|\geq2$ and 
$|Var(\beta) - Var(\alpha)|\geq2$, with $|\alpha|\geq3$ and $|\beta|\geq3$. When $x=y=1$,
we set all literals in $\alpha$ to $0$, giving us a change of measure of $5$. When $x=\neg y$, 
we remove $x$, and at least 4 variables will drop in measure, giving us $1+4\times(1-0.8039)=1.7844$.
When $x=y=0$, then set all literals in $\beta$ to $0$, giving us a change of measure of $5$. 
This gives us $\tau(5,1.7844,5)=1.3633$.

Case $C=(x \vee y \vee \alpha)$ and $C'=(x \vee \neg y \vee \beta)$.
We branch $x=y=1$ (first branch), $x=\neg y=1$ (second branch) and $x=0$ (third branch). 
For the first branch, we remove all the variables in $C$ and drop $\beta$ to a $C^1$ clause, 
for the second branch, we remove all the variables in $C'$ and drop $\alpha$ to a $C^1$ clause,
and third only $x$.

$|\alpha|=|\beta|=3$.  First branch gives $5+3\times(1-0.8039)=5.5883$. Second branch is similar, giving
$5.5883$. For the third branch, we have removed 1 variable. We take two times a debt of $0.23$, in total $0.46$ on the third branch.
This gives us $\tau(5.5883,5.5883,1+0.46)=1.3587$. The clearing of the debt
of $0.46 = 2 \times 0.23$ will be done by two 4-literal $C^2$ clauses
$(y \vee \alpha)$ and $(\neg y \vee \beta)$. First one branches
$(y \vee \alpha)$ such that the variable $y$ survives the linking
(though other variables might be linked into it); then one branches
the other clause, so that we will get the branching factor of 
$\tau(3-0.23,2-0.23)=1.3644$. \\
$|\alpha|=3,|\beta|\geq4$. First branch gives $5+4\times(1-0.8039)=5.7844$. Second branch gives
$6+3\times(1-0.8039)=6.5883$. Third branch gives $1$. We take a debt of $0.23$ for the 
third branch, giving us $\tau(5.7844,6.5883,1+0.23)=1.3530$. The debt clearance will be done
on a $C^2$ 4-literal clause, giving us $\tau(3-0.23,2-0.23)=1.3644$. \\
$|\alpha|\geq4,|\beta|\geq4$. First and second branch gives at least $6+4\times(1-0.8039)=6.7844$.
The third branch gives $1$. This gives us $\tau(6.7844,6.7844,1)=1.3510$. \\
This completes the case for 2 overlapping variables between $C$ and $C'$.

\item $|Var(C )\cap Var(C')|\geq3$. In this case, then we are able to choose $3$ variables
$x,y,z$ from $C$ and $C'$ such that we either have 
$C=(\alpha \vee \gamma)$ and $C'=(\beta \vee \gamma)$,$|\gamma|\geq3$ and
$\gamma$ contains $x,y,z$, or 
$C=(\alpha \vee x \vee y \vee z)$ and $C'=(\beta \vee x \vee y \vee \neg z)$. \\
Note that $|\alpha|\geq2$ and  $|\beta|\geq2$. We distinguish two subcases.
\begin{itemize}
\item 
Case $C=(\alpha \vee x \vee y \vee z)$ and $C'=(\beta \vee x \vee y \vee \neg z)$.
We branch $x=\neg y$ (first branch) and $x=y=0$ (second branch). 
Note that we cannot have $x=y=1$ here, else one of the clause is not satisfiable. 
The first branch removes $y$ and sets all the variables in $\alpha$, $\beta$ and $z$
to be a $C^1$ clause. The second branch removes $x,y$.

First branch gives $1+5\times(1-0.8039)=1.9805$.
Second branch gives $2$. Taking a debt of $0.54$ for the first branch, we have
$\tau(1.9805+0.54,2)=1.3609$. For the first branch, note that we are left with the
following $C^1$ clauses: $C=(\alpha \vee z)$ and $C'=(\beta \vee \neg z)$. We branch $z=1$
and $z=0$.

$|\alpha|=|\beta|=2$. In either $z=1$ ($z=0$), we remove the variables 
in $\alpha$ ($\beta$) and link up the literals in $\beta$ ($\alpha$), giving us $3\times0.8039+0.6078=3.0195$.
Then we have $\tau(3.0195-0.54,3.0195-0.54)=1.3226$. \\ 
$|\alpha|=2,|\beta|\geq3$. When $z=1$, we remove all variables in $\alpha$, giving us 
$3\times0.8039=2.4117$. When $z=0$, we remove all the variables in $\beta$, and link up
the variables in $\alpha$, giving us $4\times0.8039+0.6078=3.8234$. This gives us 
$\tau(3.8234-0.54,2.4117-0.54)=1.3182$. \\ 
$|\alpha|\geq3,|\beta|\geq3$. When $z=1$ ($z=0$), we remove all the $x$ and the variables
in $\alpha$ ($\beta$), giving us $4\times0.8039=3.2156$. This gives us a branching factor of 
$\tau(3.2156-0.54,3.2156-0.54)=1.2958$.
\item 
Case $C=(\alpha \vee \gamma)$ and $C'=(\beta \vee \gamma)$.
Then we branch $\gamma=2$, $\gamma=1$, 
and $\gamma=0$. Note also that $|\alpha|\geq2$ and $|\beta|\geq2$. 
When $\gamma=2$, all variables in $\alpha$ and $\beta$ are removed.
When $\gamma=1$, then $\gamma$, $\alpha$ and $\beta$ drops to become $C^1$
clauses. When $\gamma=0$, then all literals in $\gamma$ are assigned 0.
Note that $|\gamma|\geq3$.

$|\alpha|=|\beta|=2$. When $\gamma=2$, we remove 4 variables.
When $\gamma=1$, we link up the two literals in $\alpha$ and $\beta$,
giving us $2+3\times(1-0.8039)=2.5883$. When $\gamma=0$, we remove all
7 variables, since all literals in $\alpha$ and $\beta$ must be assigned 1. 
This gives us $\tau(4,2.5883,7)=1.3057$. \\
$|\alpha|=2,|\beta|=3$. When $\gamma=2$, we remove 5 variables.
When $\gamma=1$, we link up the two literals in $\alpha$, giving us 
$1+6\times(1-0.8039)=2.1766$. When $\gamma=0$, then all literals in
$\alpha$ are assigned 1, and the variables in $\beta$ drop their weight
from $1$ to $0.8039$, giving us a change of measure of $5+3\times(1-0.8039)=5.5883$. This gives us 
$\tau(5,2.1766,5.5883)= 1.3230$. \\
$|\alpha|=|\beta|=3$. When $\gamma=2$, we remove 6 variables. When
$\gamma=1$, we can only factor in the change of measure of $9\times(1-0.8039)=1.7649$.
When $\gamma=0$, we remove 3 variables and the variables in $\alpha$ and $\beta$
drop their weight, giving us $3+6\times(1-0.8039)=4.1766$. This gives us 
$\tau(6,1.7649,4.1766)=1.3671$. \\
$|\alpha|=3,|\beta|\geq4$. When $\gamma=2$, we remove 7 variables. When
$\gamma=1$, we can only factor in the change of measure of $10\times(1-0.8039)=1.961$.
When $\gamma=0$, we remove 3 variables and the variables in $\alpha$ and $\beta$
drop their weight, giving us $3+3\times(1-0.8039)=3.5883$. This gives us 
$\tau(7,1.961,3.5883)=1.3579$. \\
$|\alpha|\geq4,|\beta|\geq4$. When $\gamma=2$, we remove 8 variables. When
$\gamma=1$, we can only factor in the change of measure of $11\times(1-0.8039)=2.1571$.
When $\gamma=0$, we remove 3 variables and the variables in $\alpha$ and $\beta$
drop their weight, giving us $3$. This gives us 
$\tau(8,2.1571,3)=1.3592$.
\end{itemize}
This completes the case where $|Var(C) \cap Var(C')|\geq2$, for any two clause $C$ and $C'$.
\end{itemize}

From this line onwards, we do not have to worry about the increase of weights of variable, since they
are all already at weight 1, the maximum weight.

\item If there is a heavy variable $x$ in the formula that matches the subcases below, we branch it $x=1$ and $x=0$.

By the earlier lines of the algorithm, any $C^2$ clause must be at least length 5. In addition,
given any two $C^2$ clauses, we can only have $|Var(C) \cap Var(C')|\leq1$. 

\begin{itemize}
\spaceo
\item If we have $(x \vee \alpha)$, $(x \vee \beta)$ and $(\neg x \vee \gamma)$.
Note that $|\alpha|\geq4$ and $|\beta|\geq4$. We split into two cases. 

\smallskip
Case: $|\gamma|\geq5$. When $x=1$, then $\alpha$ and $\beta$
drops to a $C^1$ clause, giving us a change of measure of
$1+8\times(1-0.8039)=2.5688$. When $x=0$, we remove 
$x$ and $\gamma$ drops to a $C^1$ clause, giving us a change of measure of $1+5\times(1-0.8039)=1.9805$.
Now, this gives $\tau(2.5688,1.9805)=1.3587$.

\smallskip
Case: $|\gamma|=4$. When $x=1$, then $\alpha$ and $\beta$ drops to
$C^1$ clauses, giving us a change of measure of $1+8\times(1-0.8039)=2.5688$.
In addition, $\gamma$ becomes a $C^2$ 4-literal clause. 
We take a debt of $0.23$ for this branch.
When $x=0$, we remove
$x$ and $\gamma$ drops to a $C^1$ clause, giving us a change of measure of 
$1+4\times(1-0.8039)=1.7844$.
This gives us
$\tau(2.5688+0.23,1.7844)=1.3604$.
In the branch of $x=1$, we subsequently clear
the debt by branching the $C^2$ 4-literal clause $\gamma$ which
gives us $\tau(3-0.23,2-0.23)=1.3644$.

\item If we have $(x \vee \gamma)$, $(x \vee \alpha)$ and $(x \vee \beta)$.
Note that $|\gamma|\geq4$, $|\alpha|\geq4$ and $|\beta|\geq4$. We deal with this case by case.
When $x=1$, $\delta$, $\alpha$ and $\beta$ drops to a $C^1$ clause.
When $x=0$, we remove only $x$.

\smallskip
Case: $|\alpha|=4$, $|\beta|=4$, $|\gamma|=4$. When $x=1$, 
this gives us a change of measure of $1+12\times(1-0.8039)=3.3532$.
When $x=0$, we remove $x$ and we take a debt of $0.69$.
This gives us $\tau(3.3532,1.69)= 1.3311$. 
The follow up debt clearance will be done by the three $C^2$ 4-literal
clause, each paying back $0.23$,
giving us $\tau(3-0.23,2-0.23)=1.3644$. 

\smallskip
Case: $|\alpha|=4$, $|\beta|=4$, $|\gamma|\geq5$. When $x=1$, 
this gives us a change of measure of $1+13\times(1-0.8039)=3.5493$.
When $x=0$, we take a debt of
$0.46$. This gives us $\tau(3.5493,1.46)=1.3438$.
The debt will be cleared by the two $C^2$
4-literal clause, each paying back $0.23$,
giving us again $\tau(3-0.23,2-0.23)=1.3644$.

\smallskip
Case: $|\alpha|=4$, $|\beta|\geq5$, $|\gamma|\geq5$. When $x=1$, 
this gives us a change of measure of $1+14\times(1-0.8039)=3.7454$.
When $x=0$, we take a debt of $0.23$.
This gives us $\tau(3.7454,1.23)=1.3609$.
Again, the debt clearance is the same as above. \\

\smallskip
Case: $|\alpha|\geq5$, $|\beta|\geq5$, $|\gamma|\geq6$.
When $x=1$, this gives us a change of measure of 
$1+16\times(1-0.8039)=4.1376$. When $x=0$, we remove only $x$.
We take a debt of $0.2$ when $x=1$,
this gives us $\tau(4.1376+0.2,1)=1.3592$.
The debt of $0.2$ will be repaid by $\gamma$, which is by then a
$C^1$ clause under that branch. Choose two literals $a,b$
from $\beta$ and branch $a=\neg b$ and $a=b=0$.
The $a=\neg b$ gives us $4\times0.8039+0.6078=3.8234$.
When $a=b=0$, this gives us $2\times0.8039=1.6078$.
This gives us $\tau(3.8234-0.2,1.6078-0.2)=1.3454$.

\item Two heavy variables $x$ and $y$ appear in the same clause.
With the earlier cases not applying, for this case to happen,
$x$ and $y$ must both appear in three $C^2$ 6-literal clauses,
with a $C^2$ 6-literal clause containing
both $x$ and $y$. Let $(x \vee y \vee \alpha)$, $(x \vee \beta)$,
$(x \vee \gamma)$, $(y \vee \delta)$ and
$(y \vee \varepsilon)$ and $|\beta|=|\gamma|=|\delta|=|\varepsilon|=5$
and $|\alpha|=4$. 
We branch $x=y=1$ (first branch),
$x=1,y=0$ (2nd branch), $x=0,y=1$ (third branch)
and $x=y=0$ (fourth branch).
Note that $|Var(\beta \cup \gamma \cup \delta \cup \varepsilon)|\geq16$.
First branch, we remove $x,y$ and the variables
in $\alpha$, then $\beta$, $\gamma$, $\delta$ and $\varepsilon$
will drop in measure, giving us $6+16\times(1-0.8039)=9.1376$.
Though $\beta,\gamma,\delta,\varepsilon$ have $20$ variables
altogether, it cannot be excluded that $\beta,\gamma$ share both
one variable with each of $\delta,\varepsilon$, therefore only $16$
variables are taken into account in order to avoid double counting.
For the second branch, when 
we remove $x$, $\alpha$, $\beta$ and $\gamma$ will drop to a $C^1$ clause,
giving us a change of measure of $2+14\times(1-0.8039)=4.7454$.
The third branch is symmetric to the second, only using
$\alpha,\delta,\varepsilon$ in place of $\alpha,\beta,\gamma$.
For the fourth branch, it gives us $2$.
We take a debt of $0.0390$ on the first, second and third branch
and $0.2450$ on the fourth branch. This gives us
$\tau(9.1376+0.039, 4.7454+0.039, 4.7454+0.039, 2+0.2450)=1.3673$. \\
Debt of $0.2450$ will be cleared by $\alpha$, which is
$C^2$ 4-literal clause, giving us 
$\tau(3-0.245,2-0.245)=1.3673$.
On the other hand, the debt of $0.039$ will be cleared by branching
two disjoint $C^1$ 5-literal clauses, each paying back $0.0195$.
Choose two literals $a,b$ in $\beta$, and branch $a=\neg b$,
$a=b=0$. This gives us $\tau(3,1.5883)=1.3671$. Either $(\beta,\gamma)$
or $(\delta,\varepsilon)$ are two created $C^1$ 5-literal clauses
which are disjoint.
\end{itemize}
\item 
When the other earlier cases no longer apply, then we
brute force the remaining heavy variables.

We bound the number of heavy variables that we need to branch in this case.
Note that all the variables must have weight $1$ at this point in time. Now let $x$
be a heavy variable. The only case that is left is that $x$ appears in three 
$C^2$ 6-literal clauses, $(x \vee \alpha)$, $(x \vee \beta)$ and 
$(x \vee \delta)$. Note that $|Var(\alpha) \cup Var(\beta) \cup Var(\delta)|=15$.
For each $y \in Var(\alpha) \cup Var(\beta) \cup Var(\delta)$, $y$ can be neighbours
to at most 2 heavy variables (1 of them being $x$). Note that by the previous lines of the algorithm,
$y$ cannot be neighbours to 3 heavy variables. Since $y$ is neighbour to at most two heavy
variable and since there are 15 neighbours to $x$, they contribute a total ratio of $\frac{15}{2}$ 
(non-heavy variables to heavy variables). Therefore, for each heavy variable $x$, we have
a ratio of $1 + \frac{15}{2} = \frac{17}{2}$. The number of heavy variables among all the variables
is then $\frac{2}{17}n$. Therefore branching these heavy variables will take at most
$O(2^{\frac{2}{17}n}) = O(1.0850^n)$ time.
\end{enumerate}

\noindent
By the algorithm above, we see that all heavy variables have been dealt with,
and therefore, we can safely solve the problem in polynomial time once
that happens in Line 18 of the algorithm. Correctness of the algorithm comes from the
fact that all cases have been handled as shown in the previous section. Comparing all the branching factors that
we have computed earlier, we have the following result.

\begin{theorem}
Our G$2$XSAT algorithm runs in $O(1.3674^n)$ time.
\end{theorem}

\section{G3XSAT (Polynomial Space)}

\noindent
In this section, we give a $O(1.5687^n)$ time algorithm to solve G$3$XSAT. Unlike our 
G$2$XSAT algorithm which has alot of cases, our G$3$XSAT algorithm is simpler and deals directly with 
$C^1$, $C^2$ and $C^3$ clauses by branching and removing them in that order. 
The idea here is to take advantage of the change in measure when a $C^i$ clause drops to a $C^j$ clause, $j<i$.  \\

\subsection{Nonstandard measure for the G3XSAT algorithm}

\noindent
Let $v_i$ be a variable in $\varphi$.
Let $j$ be the lowest number such that $v_i$ appears
in a $C^j$ clause; without loss of generality,
every variable appears in a clause, as other
variables can be ignored. We define the weight $w_i$
for $v_i$ as the following for a $C^j$ clause :
\[
 w_i = 
\begin{cases}
    0.6985 ,& \text{if $j=1$;} \\
    0.875,    & \text{if $j=2$;} \\
    1,         & \text{if $j=3$.} \\
\end{cases}
\]
Like in the G$2$XSAT algorithm, the weights chosen here are optimal values given by our linear search computer program 
to bring our overall runtime for this algorithm to as low as possible. Variables in $C^1$ have weight $0.6985$, 
variables in $C^2$ have weight $0.875$, and finally variables in $C^3$ have weight 1. 
If a variable $v$ appears in clauses $C^i$ and $C^j$, where $i\neq j$ then
we assign $v$ the lower weight. Again, note that our measure $\mu = \sum_{i} w_i \leq n$.
Therefore, this gives us $O(c^\mu) \subseteq O(c^n)$.

\subsection{Algorithm}

Similar to our G$2$XSAT algorithm, we will give each line of the algorithm, followed by the
its analysis of different cases. Again, Line 1 has highest priority, followed by Line 2, etc.
We call our branch and bound algorithm G$3$XSAT(.). \\

\noindent
Input: A formula $\varphi$ \\
Output : $1$ for satisfiable $\varphi$ and $0$ for unsatisfiable $\varphi$.
\begin{enumerate}
\spaceo
\item If there are any clauses $C^i$ which is not satisfiable, return 0.
\item If there are clauses $C^i$ where there is a literal $x$ appearing $j$ times, $j>i$,
then return G$3$XSAT$(\varphi[x=0])$. 
\item If there is a $C^i$ clause $(x \vee \neg x \vee \delta)$ or $(1 \vee \delta)$,
for some literal $x$ and $\delta$, then let the new clause be
$C^{i-1} = \delta$, for $i>1$. Let $\varphi'$ be the updated formula
and return G$3$XSAT$(\varphi')$. If $i=1$, remove that clause and 
return G$3$XSAT$(\varphi[\delta=0])$.
\item If there is a $C^i$ clause $(0 \vee \delta)$, then update the $C^i$ clause as $\delta$.
Update the formula as $\varphi'$. Return G$3$XSAT($\varphi'$). 
\item If there is a $C^i$ clause, for $i>1$, with each literal in that clause appearing
exactly $i$ times, then drop the clause to a $C^1$ clause with each literal
appearing exactly once. Let $\varphi'$ be the new formula and
return G$3$XSAT$(\varphi')$. 
\item  Let $C$ be a $C^1$ clause. Then we choose any literal $x$ and $y$
and branch $x=\neg y$ and $x=y=0$. Return G$3$XSAT($\varphi[x=\neg y]$) $\vee $
G$3$XSAT($\varphi[x=y=0]$)
\item Let $C$ be a $C^2$ clause with $C=(x \vee x \vee \delta)$, for
some literal $x$. Then we can either simplify this case further or branch $x=1$ and $x=0$. 
If we simplify this case, then let $\varphi'$ bet the updated formula. Return G$3$XSAT($\varphi'$).
Else, then return G$3$XSAT($\varphi[x=1]$) $\vee$ G$3$XSAT($\varphi[x=0]$).
\item Let $C$ be a $C^2$ clause, where $|C|\geq3$. 
We deal with such clauses by choosing and branching certain variables. We choose $k$ literals
$x_1,x_2,...,x_k$ to branch $b_1,b_2,...,b_k \in\{0,1\}$. Return G$3$XSAT($\varphi[x_1=b_1]$) $\vee$
G$3$XSAT($\varphi[x_2=b_2]$) $\vee ... \vee$ G$3$XSAT($\varphi[x_k=b_k]$).
\item A $C^3$ clause $C$ containing a literal $x$ appearing more than once.
Therefore, we can have $C=(x \vee x \vee x \vee \delta)$ or $C=(x \vee x \vee \delta)$,
for some literal $x$ and some subclause $\delta$. 
If we can either simplify this further or branch the variable $x=1$ and $x=0$. If we were to simplify this case,
then let $\varphi'$ be the new updated formula. Return G$3$XSAT($\varphi'$). If we were to branch, then
return G$3$XSAT($\varphi[x=1]$) $\vee$ G$3$XSAT($\varphi[x=0]$). 
\item Let $C$ be a $C^3$ clause. Choose variables and then branch on them. Let the variables be
$x_1,x_2, ..., x_k$ taking on values $b_1,b_2, ...,b_k \in \{0,1\}$.Return 
G$3$XSAT($\varphi[x_1=b_1]$) $\vee$ G$3$XSAT($\varphi[x_2=b_2]$) $\vee ... \vee$
G$3$XSAT($\varphi[x_k=b_k]$).
\end{enumerate}

\noindent
We now show that all cases have been covered in this algorithm. Line 1 helps us to deal with any clauses that are not 
satisfiable. Line 2 helps us to remove literals appearing $j$ times
in a $C^i$ clause where $j>i$. Line 3 helps us to remove the constant 1 in $C^i$ clauses, by either removing them
(if they are satisfied) or by downgrading them to a $C^j$ clause, where $j<i$. Line 4 helps us to remove the constant 0
in the clause. After this line, no constants exist in the formula. Line 5 helps us to deal with special cases
where a literal $i$ times in a $C^i$ clause, allowing us to downgrade it to a $C^1$ clause. Lines 1 to 5 are 
simplification rules. Line 6 onwards are all branching rules. 

Line 6 helps us to remove all $C^1$ clauses by branching them. After which, no $C^1$ clauses exist in the formula. 
Line 7 helps us to deal with $C^2$ clauses having a literal appearing more than once. Line 8 helps us to deal with the 
remaining $C^2$ clauses (clause where each literal only appear once). 
Line 9 helps us to deal with $C^3$ clauses having a literal appearing more than once.
Finally, Line 10 helps us to deal with the remaining $C^3$ clauses (clause where each literal only appear once). 
This completes all the case in our algorithm. 
Note that we have covered every case in our algorithm to solve G$3$XSAT. Therefore, 
this shows the correctness of our algorithm. \\

\subsection{Analysis of Our Algorithm}

Note that when linking variables in $C^1$ or $C^2$ clauses, the measure of the surviving variable
may increase. Therefore, we will fix the following change in measure when linking in $C^1$ and $C^2$ clauses:
$p=0.6985 - (1-0.6985)=0.397$ when linking in $C^1$ clauses, and 
$q=0.875 - (1-0.875)=0.75$ when linking in $C^2$ clauses.

As shown in the previous section, Lines 1 to 5 of the algorithm are simplification rules.
We will therefore analyse the algorithm indepth from Lines 6 onwards, since they are 
branching rules. 

\begin{enumerate}
\spaceo
\setcounter{enumi}{5}
\item Let $C$ be a $C^1$ clause. Then we choose any literal $x$ and $y$
and branch $x=\neg y$ and $x=y=0$. 

This line deals with all the $C^1$ clauses. Now let $C$ be a 
$C^1$ clause. If $|C|=2$, then for the two literals $x,y$ in $C$, we
know that $x=\neg y$. Therefore, $|C|\geq3$. 
\begin{itemize}
\item $|C|=3$. Let $C=(x \vee y \vee z)$. When $x=\neg y$, we remove $z$, 
giving us a change of measure of $0.6985+p=1.0955$.
When $x=y=0$, then $z=1$. This gives us $3\times0.6898=2.0955$. 
Therefore, we have $\tau(1.0955,2.0955)=1.5687$.

\item $|C|=4$. Let $C=(x \vee y \vee z \vee w)$. When $x=\neg y$, we
remove $z,w$, giving us a change of measure of 
$2\times0.6985+p=1.794$. When $x=y=0$, then we know that $z=\neg w$.
This also gives $2\times0.6985+p=1.794$.
Therefore, we have a branching factor of $\tau(1.794,1.794)=1.4717$.

\item $|C|\geq5$. We branch $x=\neg y$ to remove at least 4 variables,
with 1 of them via linking. This gives us a change of measure of
$3\times0.6985+p=2.4925$. On the other hand, when $x=y=0$,
we only remove 2 variables, giving a change of measure of $2\times0.6985=1.397$. 
This gives us a branching factor of at most
$\tau(2.4925,1.397)=1.4431$. 
\end{itemize} 
This completes the case for $C^1$ clauses.
Note that the increase of weights due to linking variables have been factored in our 
branching factors above. 

\item Let $C$ be a $C^2$ clause with $C=(x \vee x \vee \delta)$, for
some literal $x$. 
Then we can either simplify this case further or branch $x=1$ and $x=0$.

At this point in time, no $C^1$ clauses are in the formula. Hence, all variables must
have weight at least $0.875$. We deal with the different cases below.
\begin{itemize}
\item $|C|=3$. Then $|\delta|=1$. Say $y$ is the only literal in $\delta$. Then
we must have that $y=0$ and $x=1$. No branching is required here.
\item $|C|=4$. Since Line 3 of the algorithm does not apply anymore, we know
that $|Var(\delta)|=2$. Here, we branch
$x=1$ and $x=0$. When $x=1$, we remove $y$ and $z$. Giving us a 
change of measure of $3\times0.875=2.625$. When $x=0$, $y=z=1$. This gives us
a change of measure of $3\times0.875=2.625$ as well. Therefore, we have a branching
factor of $\tau(2.625,2.625)=1.3023$. 
\item $|C|\geq5$. Note that if we have the clause $(x \vee x \vee y \vee y \vee z)$,
then we know that $z=0$. So, we can simplify this case further. Therefore, 
with Line 3 not applying, 
if $|C|\geq5$, we must have at least 3 different variables in $\delta$. Branching
$x=1$ will remove all variables, giving us a change of $4\times0.875=3.5$. On the
other hand, when $x=0$, we only remove $x$, with a change of measure of $0.875$.
This gives us $\tau(3.5,0.875)=1.4454$.
\end{itemize}
This completes the case for $C^2$ clauses containing a literal that appears twice. Here,
since we are either simplifying some cases or branching $x=1$ and $x=0$ and are not
doing any linking here, there will not be an increase in weights for any variable.

\item Let $C$ be a $C^2$ clause, where $|C|\geq3$. 
We deal with such clauses by choosing and branching certain variables. 

At this point, all literals in any $C^2$ clause must appear once.
We handle the different cases below.

\begin{itemize}
\item $|C|=3$. Let $C = (x \vee y \vee z)$. Then we negate all the literals in $C$ to 
get $C=(\neg x \vee \neg y \vee \neg z)$. Then $C$ becomes a $C^1$ 3-literal clause
and the overall measure drops.   

\item $|C|=4$. Then let $C=(x\vee y \vee z \vee w)$. We branch 
$x=\neg y$, $z=\neg w$ and $x=y,z=w$ and $y=\neg w$. For the
branch $x=\neg y$, $z=\neg w$, we get a change of measure of
$2q=1.5$. For the branch $x=y,z=w$ and $y=\neg w$,
we get a change of measure of $2\times0.875+q=2.5$. This gives
us a branching factor of $\tau(1.5,2.5)=1.4454$.

\item $|C|=5$. Let $C=(x \vee \delta)$. Then when $x=1$ (first branch), $\delta$ becomes a $C^1$ 4-literal clause.
This gives us a change of measure of $0.875+4\times(0.875-0.6985)=1.581$. When $x=0$ (second branch), $\delta$
becomes a $C^2$ 4-literal clause., giving us a change of measure of $0.875$. 
This gives us an initial branching factor of $\tau(1.581,0.875)$. Now we apply vector addition on the 1st and 
2nd branch. For the 1st branch, we remove a $C^1$ 4-literal clause, giving us a change of measure of 
$\tau(1.794,1.794)$. For the second branch, we remove a $C^2$ 4-literal clause, giving us a change 
of measure of $\tau(1.5,2.5)$. This gives us a total branching factor of 
$\tau(1.581+1.794,1.581+1.794,0.875+1.5,0.875+2.5)=1.5686$.

\item $|C|=6$. We choose two literals $x,y$ and branch $x=y=1$ (1st branch), $x=\neg y$ (2nd branch) and 
$x=y=0$ (3rd branch). When $x=y=1$, we remove all variables, giving us a change of measure of 
$6\times0.875=5.25$. When $x=\neg y$, we remove $x$ and the remaining clause drops to a $C^1$ 4-literal
clause, giving us a change of measure of $q+4\times(0.8756-0.6985)=1.456$. When $x=y=0$,
we remove $x,y$, giving us $2\times0.875=1.75$.
This gives us an initial branching factor of $\tau(5.25,1.456,1.75)$. We apply vector addition to the 
3rd branch. For the 2nd branch, we get a $C^2$ 4-literal clause, which gives us a branching
factor of $\tau(1.5,2.5)$ from $|C|=4$. Applying branching vector addition to the
2nd branch, we get the following branching factor 
$\tau(5.25,1.456,1.75+1.5,1.75+2.5)=1.5641$.

\item $|C|\geq7$. We choose two literals $x,y$ and branch $x=y=1$, $x=\neg y$ and $x=y=0$.
When $x=y=1$, we remove all 7 variables, giving us a change of measure of $7\times0.875=6.125$. 
When $x=\neg y$, then the clause drops to a $C^1$ 5-literal clause, giving us a change of measure
of $q+5\times(0.875-0.6985)=1.6325$. When $x=y=0$, we remove $x,y$, 
giving us a change of measure of $2\times0.875=1.75$. This gives us a branching factor of 
at most $\tau(6.125,1.6325,1.75)=1.5669$.
\end{itemize}
This completes the case for all $C^2$ clauses. Note that for the above cases where we perform 
linking of variables, the increase in weights for that surviving variable has been factored into our branching factors.

\item A $C^3$ clause $C$ containing a literal $x$ appearing more than once.
Therefore, we can have $C=(x \vee x \vee x \vee \delta)$ or $C=(x \vee x \vee \delta)$,
for some literal $x$ and some subclause $\delta$. 
If we can either simplify this further or branch the variable $x=1$ and $x=0$.

At this point onwards, all the variables in the formula must have weight 1.
If $C=(x \vee x \vee x \vee \delta)$, then branch $x=1$ and $x=0$.
\begin{itemize}
\item $|\delta|\leq2$, then $x=1$.
\item $|\delta|\geq3$. Note that if $|\delta|=3$ and since Line 3 of the algorithm does not apply anymore, we know that
there must be at least two different variables in $\delta$. Therefore, branching $x=1$ removes all 3 variables, while
$x=0$ only removes 1. This gives us a branching factor of $\tau(3,1)=1.4656$.
\end{itemize}

Now, we deal with the case that $C=(x \vee x \vee \delta)$ and there are no literals $y$ appearing three times.
\begin{itemize}
\item $|\delta|=2$. Then $\delta$ must have two different variables. If not, then the case $(x \vee x \vee y \vee y)$ is not
satisfiable. For this case, $x=1$. No branching is involved here.
\item $|\delta|=3$. If $|Var(\delta)|=2$, then we have $(x \vee x \vee y \vee y \vee z)$. Then $x=\neg y$. 
Else $|Var(\delta)|=3$. Here, we branch $x=1$ and $x=0$.
When $x=1$, we remove $x$, while the clause drops to a $C^1$ clause, 
giving us a change of measure of $1+2\times(1-0.6985)=1.603$.
When $x=0$, we remove the variable $x$ with weight 1, while the rest of the literals in $\delta$ be assigned $1$.
This gives us a branching factor of $\tau(1.603,4)=1.3037$.
\item $|\delta|=4$. Then $\delta$ cannot contain only 2 different variables. Else, we will have a case of 
$C=(x \vee x \vee y \vee y \vee z \vee z)$, which is not satisfiable. Therefore, $\delta$ must contain at least
3 different variables. Now if $\delta = (y \vee y \vee z \vee w)$. Then we must have that $x=\neg y$. No
branching is required here. If $\delta = (y \vee u \vee z \vee w)$, with 4 different variables in $\delta$, then
we branch $x=1$ and $x=0$. When $x=1$, we remove $x$ and it drops to a $C^1$ 4-literal clause.
This gives us a change of measure of $1+4\times(1-0.6985)=2.206$. Now $x=0$, we have a change
of measure of $1$. This gives us an initial branching factor of $\tau(2.206,1)$. For the first branch,
$\delta$ becomes a $C^1$ 4-literal clause, giving us a branching factor of $\tau(1.794,1.794)$. This gives us 
a branching factor of $\tau(2.206+1.794,2.206+1.794,1)=1.5437$.

\item $|\delta|=5$. If $\delta = (y \vee y \vee z \vee z \vee w)$, then we know that $w=1$. Hence, no branching
is involved here. If $\delta = (y \vee y \vee z \vee w \vee u)$, then we branch $x=1$ and $x=0$. When $x=1$,
then $\delta$ drops to a $C^1$ clause. In this case, $y$ must then be set to $0$. This gives us a change of measure of
$2+3\times(1-0.6985)=2.9045$. On the other hand, when $x=0$, we have a change of measure of $1$.
This gives us a branching factor of $\tau(2.9045,1)=1.4763$. If $\delta = (y \vee z \vee w \vee u \vee v)$,
then we again branch $x=1$ and $x=0$. When $x=1$, we have a change of measure of 
$1+5\times(1-0.6985)=2.5075$. When $x=0$, we have a change of measure of $1$. This gives us a branching
factor of $\tau(2.5075,1)=1.5279$.

\item $|\delta|\geq6$. If $|Var(\delta)|=3$, then we must have $\delta=(y \vee y \vee z \vee z \vee w \vee w)$.
This case is not satisfiable. $|Var(\delta)|=4$, then we must either have $\delta=(y \vee y \vee z \vee z \vee w \vee u)$ (i),
$\delta=(y \vee y \vee z \vee z \vee w \vee w \vee u)$ (ii) or $\delta=(y \vee y \vee z \vee z \vee w \vee w \vee u \vee u)$ (iii).
For (i), we have $w=\neg u$. For (ii), we must have $u=1$ and (iii) is not satisfiable.
No branching is required here. Finally, if $|Var(\delta)|\geq5$, then we branch $x=1$ and $x=0$.
When $x=1$, then we have a change of measure of $1+5\times(1-0.6985)=2.5075$. When $x=0$, we have a change of measure
of $1$. This gives us a branching factor of $\tau(2.5075,1)=1.5279$. 
\end{itemize}

This completes the case for a $C^3$ clause containing a literal $x$ appearing more than once. For this line onwards,
we do not have to worry about increase in weights of variables due to linking since every variable has weight 1 now.

\item Let $C$ be a $C^3$ clause. Choose $x$ in $C$ and branch $x=1$
and $x=0$.

At this point, all literals in $C$ must appear once. We consider all cases here
with $|C|\geq4$.

\begin{itemize}
\item $|C|=4$. Let $C=(x \vee y \vee z \vee w)$. Then let $C=(\neg x \vee \neg y \vee \neg z \vee \neg w)$.
Then $C$ becomes a $C^1$ 4-literal clause and the overall measure drops.
\item $|C|=5$. Let $C=(x \vee y \vee z \vee w \vee u)$. Then let 
$C=(\neg x \vee \neg y \vee \neg z \vee \neg w \vee \neg u)$. Then $C$ becomes a $C^2$ clause and the
overall measure drops.
\item $|C|=6$. Let $C=(x \vee \delta)$, for some literal $x$ and subclause $\delta$. When $x=1$,
$\delta$ becomes a $C^2$ clause, this gives us $1+5\times(1-0.875)=1.625$. On the other hand, 
when $x=0$, $\delta$ is a $C^3$ 5-literal clause, which we can downgrade to a $C^2$ 5-literal clause by
negating all the literals, giving us a change of measure of $1+5\times(1-0.875)=1.625$. This gives us a
branching factor of $\tau(1.625,1.625)=1.5320$.

\item $|C|=7$. Let $C=(x \vee y \vee \delta)$. We branch $x=y=1$, $x=\neg y$ and $x=y=0$.
When $x=y=1$, $\delta$ drops to a $C^1$ clause, giving us $2+5\times(1-0.6985)=3.5075$.
When $x=\neg y$, we remove $x$ and $\delta$ drops to a $C^2$ clause, giving us $1+5\times(1-0.875)=1.625$.
When $x=y=0$, $\delta$ becomes a $C^3$ 5-literal clause, which we can downgrade to a $C^2$ 5-literal clause,
giving us $2+5\times(1-0.875)=2.625$. This gives us a branching factor of $\tau(3.5075,1.625,2.625)=1.5647$.

\item $|C|=8$. Let $C=(x \vee y \vee \delta)$. We branch $x=y=1$ (first branch), $x=\neg y$ (second branch) 
and $x=y=0$ (third branch). When $x=y=1$, $\delta$ drops to a $C^1$ clause, giving us $2+6\times(1-0.6985)=3.809$.
When $x=\neg y$, we remove $x$ and $\delta$ drops to a $C^2$ clause, giving us $1+6\times(1-0.875)=1.75$.
When $x=y=0$, we remove $x,y$, giving us a change of measure of $2$. This gives us an initial branching factor of
$\tau(3.809,1.75,2)$. We apply vector addition on the 1st branch, where $\delta$ is a $C^1$ 6-literal clause. The
branching factor of a $C^1$ 6-literal clause is $\tau(5\times0.6985-(1-0.6985),2\times0.6985)=\tau(3.191,1.397)$.
This gives us a branching factor of $\tau(3.809+3.191,3.809+1.397,1.75,2)=1.5687$.

\item $|C|\geq9$. Let $C=(x \vee y \vee \delta)$. We branch $x=y=1$, $x=\neg y$ and $x=y=0$.
When $x=y=1$, $\delta$ drops to a $C^1$ clause, giving us $2+7\times(1-0.6985)=4.1105$.
When $x=\neg y$, we remove $x$ and $\delta$ drops to a $C^2$ clause, giving us $1+7\times(1-0.875)=1.875$.
When $x=y=0$, we remove only $x,y$, a change of measure of $2$. 
This gives us a branching factor of at most $\tau(4.1105,1.875,2)=1.5642$.
\end{itemize}
\end{enumerate}

\noindent
Comparing all the branching factors in the analysis above,
we have the following result: 

\begin{theorem}
The algorithm that solves G$3$XSAT runs in $O(1.5687^n)$ time.
\end{theorem}

\section{G4XSAT (Polynomial Space)}

In this section, we present an $O(1.6545^n)$ time algorithm to solve G$4$XSAT. 
Our algorithm for G$4$XSAT is similar to our algorithm to solve
G$3$XSAT and extends it. The weights used in the nonstandard measure
is different though. \\

\subsection{Nonstandard measure for the G4XSAT algorithm}

\noindent
Let $v_i$ be a variable in $\varphi$. Let $j$ be the lowest number such that $v_i$ appears
in a $C^j$ clause; without loss of generality, every variable appears in a clause, as other
variables can be ignored. We define the weight $w_i$ for $v_i$ as the following
for a $C^j$ clause :
\[
 w_i = 
\begin{cases}
    0.6464 ,& \text{if $j=1$;} \\
    0.8376, & \text{if $j=2$;} \\
    0.9412, & \text{if $j=3$;} \\
    1, & \text{if $j=4$.}
\end{cases}
\]
Similar to our G$3$XSAT algorithm, the weights chosen here are optimal values
given by our linear search computer program 
to bring our overall runtime for this algorithm to as low as possible. Variables in $C^1$ have weight 0.6464, 
variables in $C^2$ have weight 0.8376, variables in $C^3$ have weight 0.9412, and finally variables in $C^4$ have
weight 1. 

\subsection{Algorithm}

\noindent
Similar to the earlier algorithms, Line 1 has highest priority, followed by Line 2, etc.
We call our recursive algorithm G$4$XSAT(.). \\

\noindent
Algorithm: G$4$XSAT. \\
Input: A CNF formula $\varphi$. \\
Output: $1$ for satisfiable $\varphi$ and $0$ for unsatisfiable $\varphi$.

\begin{enumerate}
\spaceo
\item If any clause is not satisfiable, then return 0. If there are 
no clauses left, return 1. 

\item If there are clauses $C^i$ where there is a literal $x$ appearing $j$
times, $j>i$, then return G$4$XSAT$(\varphi[x=0])$.

\item If there is a $C^i$ clause $(x \vee \neg x \vee \delta)$ or
$(1 \vee \delta)$,
for some literal $x$ and $\delta$, then let the new clause be
$C^{i-1} = \delta$, for $i>1$. Let $\varphi'$ be the updated formula
and return G$4$XSAT$(\varphi')$. If $i=1$, remove that clause and 
return G$4$XSAT$(\varphi[\delta=0])$.  

\item If there is a $C^i$ clause $(0 \vee \delta)$, then let
the new clause be $\delta$. Update it as $\varphi'$ and
return G$4$XSAT$(\varphi')$. 

\item If there is a $C^i$ clause, for $i>1$, with each literal in that clause appearing
exactly $i$ times, then drop the clause to a $C^1$ clause with each literal
appearing exactly once. If there is a $C^4$ clause $C$ with each literal $x$ appearing
$2\times j$ times, for some $j$, then drop $C$ to a $C^2$ clause with each literal
$x$ appearing $j$ times. Let $\varphi'$ be the new formula and
return G$4$XSAT$(\varphi')$.  

\item If there is a $C^1$ clause $C$, then we choose two literals $x,y$
and branch $x=\neg y$ and $x=y=0$. Then return 
G$4$XSAT($\varphi[x=\neg y]$) $\vee$ G$4$XSAT($\varphi[x=y=0]$). 

\item Let $C$ be a $C^2$ clause with $C=(x \vee x \vee \delta)$. We branch $x=1$
and $x=0$. Return G$4$XSAT($\varphi[x=1]$) $\vee$ G$4$XSAT($\varphi[x=0]$).

\item Let $C$ be a $C^2$ clause. We choose $k$ literals
$x_1,x_2,...,x_k$ to branch $b_1,b_2,...,b_k \in\{0,1\}$. Return 
G$4$XSAT($\varphi[x_1=b_1]$) $\vee$ G$4$XSAT($\varphi[x_2=b_2]$)
$\vee ... \vee $ G$4$XSAT($\varphi[x_k=b_k]$). 

\item A $C^3$ clause $C$ containing a literal $x$ appearing more than once.
Then $C=(x \vee x \vee x \vee \delta)$ or $C=(x \vee x \vee \delta)$,
for some literal $x$ and some subclause $\delta$. We will either simplify 
the cases further or branch the variable $x=1$ and $x=0$. If we simplify, let $\varphi'$
be the updated formula. Return G$4$XSAT($\varphi'$). If we branch $x$, then
return G$4$XSAT($\varphi[x=1]$) $\vee$ G$4$XSAT($\varphi[x=0]$). 

\item Let $C$ be a $C^3$ clause. We will either simplify the cases or applying
branching here. If we were to simplify the case, we update the formula as $\varphi'$.
Return G$4$XSAT($\varphi'$). Else if we are branching, then we choose $k$ literals
$x_1,x_2,...,x_k$ to branch $b_1,b_2,...,b_k \in\{0,1\}$. Return 
G$4$XSAT($\varphi[x_1=b_1]$) $\vee$ G$4$XSAT($\varphi[x_2=b_2]$)
$\vee ... \vee $ G$4$XSAT($\varphi[x_k=b_k]$). 

\item  A $C^4$ clause $C$ containing a literal $x$ appearing more than once. We
can either simplify the cases further or we branch $x=1$ and $x=0$. If we simplify the
case, then update the formula as $\varphi'$ and return G$4$XSAT($\varphi')$. 
Else we branch $x=1$ and $x=0$ and return 
Return G$4$XSAT($\varphi[x=1]$) $\vee$ G$4$XSAT($\varphi[x=0]$).

\item Let $C$ be a $C^4$ clause. We will either simplify the cases or applying
branching here. If we were to simplify the case, we update the formula as $\varphi'$.
Return G$4$XSAT($\varphi'$). Else if we are branching, then we choose $k$ literals
$x_1,x_2,...,x_k$ to branch $b_1,b_2,...,b_k \in\{0,1\}$. Return 
G$4$XSAT($\varphi[x_1=b_1]$) $\vee$ G$4$XSAT($\varphi[x_2=b_2]$)
$\vee ... \vee $ G$4$XSAT($\varphi[x_k=b_k]$). 
\end{enumerate}

\noindent
Again, we show that we have covered all the cases in our algorithm. In Line 1, if there
is any clause that is not satisfiable, then we return 0. In Line 2, if there are any literals 
appearing $j$ times in a $C^i$ clause, $j>i$, then we can set $x=0$ since we are not
allowed to over satisfy. In Line 3 of the algorithm, if there are constants $1$ in the formula,
we can either downgrade the clause or remove the clause, depending whether the clause is satisfied
or not. In Line 4 of the algorithm, we remove the constants $0$ in the formula. After Line 4, no
more constants exist in the formula. In Line 5, we deal with special cases where every literal
in the clause appears $i$ times in a $C^i$ clause, allowing us to downgrade it to a $C^1$ clause. 
Lines 1 to 5 are simplification rules.

From Line 6 onwards, they are branching rules which contribute to the exponential time growth of
the algorithm. In Line 6 of the algorithm, we branch and deal with all $C^1$ clauses. In Line 7 of the algorithm,
we deal with $C^2$ clauses containing literals $x$ appearing twice. In Line 8 of the algorithm, we remove
the remaining $C^2$ clauses by branching them (now all $C^2$ clauses have literals in it appearing only once). 
After which, no more $C^2$ clauses exist in the formula. 
In Line 9 of the algorithm, we deal with $C^3$ clauses with literals appearing more than once. In Line 10, we
remove the remaining $C^3$ clauses, where every literal appearing in the $C^3$ clauses only appears once.
In Line 11, we deal with $C^4$ clauses with literals appearing more than once in it. Finally, in Line 12, we remove
all $C^4$ clauses, where every literal in the clause appears only one. Therefore, we have covered all cases and
hence the algorithm is correct. Note that our G$4$XSAT algorithm is just an extension of our G$3$XSAT algorithm,
by handling $C^4$ clauses. \\

\subsection{Analysis of Our Algorithm}
In this section, we analyse the time complexity of our G$4$XSAT algorithm.
Similar to our previous algorithm, whenever we link variables in $C^i$ clauses, where $i<4$, we need to factor in
the increase of measure because the surviving variable may now appear in a $C^4$ clause. 
We have to factor in an increase of measure of  
$p=0.6464 - (1-0.6464)=0.2928$ when linking in $C^1$ clauses, and 
$q=0.8376 - (1-0.8376)=0.6752$ when linking in $C^2$ clauses, and 
$r =0.9412 - (1-0.9412)=0.8824$ when linking in $C^3$ clauses.

Since Lines 1 to 5 are simplification rules, they will only take polynomial time. Hence, we will analyse
from Lines 6 onwards since they are all branching rules. 

\begin{enumerate}
\spaceo
\setcounter{enumi}{5}

\item If there is a $C^1$ clause $C$, then we choose two literals $x,y$
and branch $x=\neg y$ and $x=y=0$.
\begin{itemize}
\item $|C|=3$. Then let $C=(x \vee y \vee z)$. When $x=\neg y$,
we remove $z$, giving us a change of measure of 
$p+0.6464=0.9392$. When $x=y=0$, then $z=1$,
giving us a change of measure of $3\times0.6464=1.9392$.
This gives us a branching factor of $\tau(0.9392,1.9392)=1.6544$.
\item $|C|=4$. Then let $C=(x \vee y \vee z \vee w)$. Then 
when $x=\neg y$, then $z=w=0$. This gives a change of 
$2\times0.6464+p=1.5856$. When $x=y=0$,
then $z=\neg w$. This gives us a change of measure of 
$2\times0.6464+p=1.5856$. Therefore,
we have a branching factor of $\tau(1.5856,1.5856)=1.5483$.
\item $|C|\geq5$. Let $C=(x \vee y \vee z \vee w \vee u)$. 
When $x=\neg y$, then $z=w=u=0$. This gives us a change of measure
of $3\times0.6464+p=2.232$. On the other hand,
when $x=y=0$, we only remove the 2 variables, giving
us a change of measure of $2\times0.6464=1.2928$. This gives us a branching
factor of $\tau(2.232,1.2928)=1.4970$. 
\end{itemize}

This completes the case for all $C^1$ clause. 
Note that the increase of weights due to linking variables have been factored in our 
branching factors above. \\

\item Let $C$ be a $C^2$ clause with $C=(x \vee x \vee \delta)$. 

From this point on, all variables have at least weight $0.8376$.
Then we can either simplify this case further or branch $x=1$ and $x=0$.

\begin{itemize}
\item $|\delta|=1$. Then $x=1$. No branching is involved here.
\item $|\delta|=2$. Then if $\delta = (y \vee y)$,then this case
would have already been handled by Line 5 of the algorithm. 
If $\delta= (y \vee z)$, for some literal $y,z$, then $y=z$. 
\item $|\delta|=3$. Then if $\delta=(y \vee y \vee z)$,then $z=0$. 
If $|Var(\delta)|\geq3$, then when $x=1$, we remove all 4
variables, giving us a change of measure of $4\times0.8376=3.3504$. When
$x=0$, we only have a change of measure of $0.8376$. This gives us a 
branching factor of $\tau(3.3504,0.8376)= 1.4693$.
\item $|\delta|\geq4$. Since Line 5 of the algorithm does not apply anymore,
then we know that $|Var(\delta)|\geq3$.
Branching $x=1$ and $x=0$ will give us a branching factor of at most
$\tau(3.3504,0.8376)= 1.4693$.
\end{itemize}

This concludes the case for $C^2$ clauses having a literal appearing twice in it. Here, since
we are only simplifying the cases or branching $x=1$ and $x=0$, there will be no increase in weights
of variables. \\

\item Let $C$ be a $C^2$ clause. At this point in time, there are no literals in $C$ that appears twice.
We handle this case by case again.

\begin{itemize}
\item $|C|=3$. Let $C=(x \vee y \vee z)$. Then let $C=(\neg x \vee \neg y \vee \neg z)$
be the new clause obtained by negating all the literals in it. In this case, $C$ becomes a
$C^1$ 3-literal clause and the overall measure decreases. No branching is needed for this case. 

\item $|C|=4$. Let $C=(x \vee y \vee z \vee w)$. Then we branch $x=\neg y$, $z=\neg w$ and
$x=y,z=w$ and then $x=\neg z$. The first branch gives us $2q=1.3504$.
The second branch gives $2\times0.8376+q=2.3504$. This gives a branching factor of
$\tau(1.3504,2.3504)=1.4690$. 

\item $|C|=5$. Let $C=(x \vee\delta)$. We branch $x=1$ (first branch) and $x=0$ (second branch).
When $x=1$, we have a change of measure of $0.8376+4\times(0.8376-0.6464)=1.6024$.
When $x=0$, we remove $x$, giving us a change of measure of $0.8376$. This gives us an initial
branching factor of $\tau(1.6024,0.8376)$.
For the first branch, $\delta$ becomes a $C^1$ 4-literal clause. This gives us a branching factor of 
$\tau(1.5856,1.5856)$. For the second branch, $\delta$ becomes a $C^2$ 4-literal clause, giving us a branching
factor of $\tau(1.3504,2.3504)$. Applying branching vector addition to the first and second branch respectively,
this gives us $\tau(1.6024+1.5856,1.6024+1.5856,0.8376+1.3504,0.8376+2.3504)=1.6157$.

\item $|C|\geq6$. Let $C=(x \vee y \vee \delta)$. We branch $x=y=1$ (first branch), 
$x=\neg y$ (second branch), $x=y=0$ (third branch). When $x=y=1$,
then we remove at least 6 variables having weight $0.8376$, this gives us a change of measure of 
$6\times0.8376=5.0256$. When $x=\neg y$, we remove $x$ and $\delta$ drops to a $C^1$ 
clause of at least 4-literal, giving us a change of measure of $0.8376+4\times(0.8376-0.6464)-(1-0.8376)=1.44$.
Finally, when $x=y=0$, we remove 2 variables of weight $0.8376$, giving us a change
of measure of $2\times0.8376=1.6752$. This gives us a branching factor of 
$\tau(5.0256,1.44,1.6752)=1.6493$. 
\end{itemize}
This completes the case for all $C^2$ clauses. The increase in weights of variable due to linking variables
have been factored into our branching factors above. \\

\item A $C^3$ clause $C$ containing a literal $x$ appearing more than once.
Therefore, we can have $C=(x \vee x \vee x \vee \delta)$ or $C=(x \vee x \vee \delta)$,
for some literal $x$ and some subclause $\delta$. If we can either simplify this further or branch the variable $x=1$ and $x=0$.

At this point onwards, all the variables in the formula must have weight at least $0.9412$.
If $C=(x \vee x \vee x \vee \delta)$, for some literal $x$, then we split it into different cases.

\begin{itemize}
\item $|\delta| \leq 2$. Then $x=1$, no branching involved here.
\item $|\delta|\geq3$. Since Line 5 does not run anymore, we know that $\delta$ contains at least two other variables.
Therefore, branching $x=1$, will remove at least 3 variables in $C$, giving us a change of measure of 
$3\times0.9412=2.8236$. On the other hand, when $x=0$, we have a change of measure of $0.9412$. 
This gives us a branching factor of $\tau(2.8236,0.9412)= 1.5010$. 
\end{itemize}

Now, if literals in $C$ appear at most twice, then let $C=(x \vee x \vee \delta)$.

\begin{itemize}
\item $|\delta|\leq 2$. Now If $|\delta|=1$, then all literals must be assigned $1$. Now, if $|\delta|=2$
and $\delta=(y \vee y)$, then this is not satisfiable. Now if we have 
$\delta=(y \vee z)$, then we know that $y=\neg z$ and $x=1$. 

\item $|\delta|=3$. If $\delta=(y \vee y \vee z)$, then $z=1$. 
If $\delta=(y \vee z \vee w)$, then we branch $x=1$ and $x=0$.
When $x=1$, $\delta$ drops to a $C^1$ 3-literal clause, giving us a change of measure
of $0.9412+3\times(0.9412-0.6464)=1.8256$. When $x=0$, then $y=z=w=1$,
giving us a change of measure of $4\times0.9412=3.7648$. This gives us a branching factor of
$\tau(1.8256,3.7648)=1.2959$.

\item $|\delta|\geq4$. If $|Var(\delta)|=2$, then $C$ is not satisfiable. 
If $|Var(\delta)|=3$, then (i) $\delta = (y \vee y \vee z \vee w)$  or 
(ii) $\delta = (y \vee y \vee z \vee z \vee w)$ . For (ii), $w=1$ and
no branching is involved here. For (i), when $x=1$, 
then $y=0$ and $z=\neg w$. When $x=0$, then $y=1$ and $z=\neg w$. 
In either case, we can simplify this case by setting $x=\neg y$. Finally,
if $|Var(\delta)|\geq4$, then we branch $x=1$ and $x=0$. When $x=1$,
$\delta$ drops to a $C^1$ clause of at least 4-literal, giving us a change of measure of 
$0.9412+4\times(0.9412-0.6464)=2.1204$. When $x=0$, we have a change of measure of $0.9412$.
This gives us a branching factor of $\tau(2.1204,0.9412)=1.6136$.
\end{itemize}

This completes the case of a literal $x$ appearing in a $C^3$ clause more than once. Here, since
we are only simplifying the cases or branching $x=1$ and $x=0$, there will be no increase in weights
of variables. \\

\item Let $C$ be a $C^3$ clause.
At this point, every literal appearing in a $C^3$ clause can only appear once.

\begin{itemize}
\item $|C|=3$. Assign all literals $1$.
\item $|C|=4$. Let $C=(x \vee y \vee z \vee w)$. Then let 
$C=(\neg x \vee \neg y \vee \neg z \vee \neg w)$. $C$ becomes a 
$C^1$ 4-literal clause and the overall measure drops.
\item $|C|=5$. Let $C=(x \vee y \vee z \vee w \vee u)$. Then let 
$C=(\neg x \vee \neg y \vee \neg z \vee \neg w \vee \neg u)$. Then
$C$ becomes a $C^2$ clause and the overall measure drops.

\item $|C|=6$. Let $C=(x \vee \delta)$. We branch $x=1$ and $x=0$.
When $x=1$, we remove $x$ and $\delta$ drops to a $C^2$ 5-literal clause, giving us
a change of measure of $0.9412+5\times(0.9412-0.8376)=1.4592$. When $x=0$,
we remove $x$ and $\delta$ becomes a $C^3$ 5-literal clause, where we can 
drop it to a $C^2$ 5-literal clause by negating all the literals, giving us a change of measure of
$0.9412+5\times(0.9412-0.8376)=1.4592$. This gives $\tau(1.4592,1.4592)=1.6081$.

\item $|C|=7$. Let $C=(x \vee y \vee \delta)$. We branch $x=y=1$ , $x=\neg y$ and
$x=y=0$. When $x=y=1$, we remove $x,y$, with $\delta$ becoming a $C^1$ 5-literal clause,
giving us a change of measure of $2\times0.9412+5\times(0.9412-0.6464)=3.3564$. When 
$x=\neg y$, we remove $x$ via linking and $\delta$ drops to a $C^2$ clause, giving us 
$r+5\times(0.9412-0.8376)=1.4004$. When $x=y=0$, $\delta$ drops to a
$C^3$ 5-literal clause, where it can be further dropped to a $C^2$ 5-literal clause, giving us
$2\times0.9412+5\times(0.9412-0.8376)=2.4004$. This gives us 
$\tau(3.3564,1.4004,2.4004)=1.6363$.

\item $|C|=8$. Let $C=(x \vee y \vee \delta)$. We branch $x=y=1$ (first branch), $x=\neg y$ (second branch)
and $x=y=0$ (third branch). Now when $x=y=1$, $\delta$ drops to a $C^1$ clause, giving us 
$2\times0.9412+6\times(0.9412-0.6464)=3.6512$. When $x=\neg y$, we remove $x$ via linking and $\delta$ drops to a
$C^2$ clause, giving us $r+6\times(0.9412-0.8376)=1.504$. When $x=y=0$, we remove $x,y$,
giving us $2\times0.9412=1.8824$. This gives us $\tau(3.6512,1.504,1.8824)$. On the third branch, $\delta$
becomes a $C^3$ 6-literal clause, with branching factor of $\tau(1.4592,1.4592)$. Applying vector addition on the
third branch, the sum of these vectors gives us a branching factor of $\tau(3.6512,1.504,1.8824+1.4592,1.8824+1.4592)=1.6544$.

\item $|C|\geq9$. Let $C=(x \vee y \vee \delta)$. We branch $x=y=1$, $x=\neg y$ and $x=y=0$. 
When $x=y=1$, we remove $x,y$ and $\delta$ drops to a $C^1$ 7-literal clause, giving us a change of measure of 
$2\times0.9412+7\times(0.9412-0.6464)=3.946$. When $x=\neg y$, we remove $x$ via linking and we are left
with a $C^2$ 7-literal clause, giving us a change of measure of $r+7\times(0.9412-0.8376)=1.6076$.
Finally, when $x=y=0$, we remove 2 variables, giving us a change of measure of $2\times0.9412=1.8824$.
This gives us a branching factor of at most $\tau(3.946,1.6076,1.8824)=1.6301$
\end{itemize}

This completes the case for all $C^3$ clauses. The increase in weights due to linking variables have
been factored into our branching factors above. \\

\item A $C^4$ clause containing a literal $x$ appearing at least twice in it.

At this point, all variables in the formula have weight 1.
If $C=(x \vee x \vee x \vee x \vee \delta)$, then $|\delta|\geq1$.
Branch $x=1$ and $x=0$. When $x=1$, we will remove another variable, giving
us a change of measure of 2. On the other hand, when $x=0$, we remove
only 1 variable. This gives us a branching factor of at most $\tau(2,1)=1.6181$.

Now, now suppose that a literal only appears at most three times in $C$, then
let $C=(x \vee x \vee x \vee \delta)$, for some literal $x$. 

\begin{itemize}
\item $|\delta|=1$. Assign all literals in $C$ to be $1$.
\item $|\delta|=2$. Then $|Var(\delta)|=2$. Let $y,z$ be the two literals
in $\delta$. Then $x=1$ and $y=\neg z$.
\item $|\delta|=3$. If $\delta=(y \vee y \vee y)$, then this case is not satisfiable. 
If $\delta=(y \vee y \vee z)$, then $x=z=1$ and $y=0$.  If $\delta =(y \vee z \vee w)$, then 
$x=1$ as $x$ cannot be $0$. 
\item $|\delta|=4$. If $\delta = (y \vee y \vee y \vee z)$, then $z=1$ and $x=\neg y$. If 
$\delta=(y \vee y \vee z \vee z)$, then $y=z=1$ and $x=0$. If $\delta=(y \vee y  \vee z \vee w)$,
or in general $|Var(\delta)|\geq3$, then we branch $x=1$ and $x=0$. 
When $x=1$, we remove $x$, and then $\delta$ drops to a $C^1$ clause,
this gives $1+3\times(1-0.6464)=2.0608$. On the other hand, when $x=0$, we remove $x$.
This gives us a branching factor of $\tau(2.0608,1)=1.6053$.
\item $|\delta|\geq5$. If $\delta$ has only two different variables, then the only cases we have is either 
$\delta = (y \vee y \vee y \vee z \vee z)$ or $(y \vee y \vee y \vee z \vee z \vee z)$. For
both cases, $C$ is not satisfiable. Therefore, there must be at least 3 different variables in
$\delta$. Therefore, branching $x=1$ and $x=0$ will give us again the branching factor
$\tau(1+3\times(1-0.6464),1)=1.6053$.
\end{itemize}

This completes the case of a literal appearing at three times in $C$.
Now if there is a literal $x$ appearing twice in $C$, then let $C=(x \vee x \vee \delta)$.
Note that if every literal appears twice in $C$, then $C$ can be downgraded to a 
$C^2$ clause with the same literal appearing exactly once. Such cases will then
be handled by earlier lines of the algorithm.

\begin{itemize}
\item $|\delta|=2$. Assign all literals in $C$ to be $1$.
\item $|\delta|=3$. If $\delta =(y \vee y \vee z)$, then $x=y=1$ and $z=0$. 
If $\delta=(y \vee z \vee w)$, then $x=1$. 
\item $|\delta|=4$. If $\delta=(y \vee y \vee z \vee z)$, then we downgrade $C$ into 
a $C^2$ clause. If $\delta = (y \vee y \vee z \vee w)$, then $z=w$.
If $\delta=(y \vee z \vee w \vee u)$, then branching $x=1$ will remove $x$,
and cause the clause to drop to a $C^2$ 4-literal clause, giving us a change of measure of 
$1+4\times(1-0.8376)=1.6496$. On the other hand, when $x=0$, then the clause drops to a 
4-literal clause, where all literals must be assigned $1$ then. This gives us a branching factor 
of $\tau(1.6496,5)=1.2591$. 
\item $|\delta|=5$. If $\delta=(y \vee y \vee z \vee z \vee w)$, then $w=0$.
If $\delta=(y \vee y \vee z \vee w \vee u)$, then we branch $x=1$ and $x=0$.
When $x=1$, $\delta$ drops to a $C^2$ clause, giving us a change of measure of 
$1+4\times(1-0.8376)=1.6496$. When $x=0$, then $y=1$, this gives us a change of
measure of at least 2. Therefore, we have $\tau(1.6496,2)=1.4639$. Now if 
$|Var(\delta)|=5$, then we branch $x=1$ and $x=0$. This gives us 
$1+5\times(1-0.8376)=1.812$ for $x=1$. When $x=0$, $\delta$ is a $C^4$ 5-literal
clause. We can negate all the literals in $\delta$, and hence $\delta$ drops to a $C^1$
5-literal clause, giving us a total change of measure of $1+5\times(1-0.6464)=2.768$.
Therefore, we have $\tau(1.812,2.768)=1.3599$. 

\item $|\delta|\geq6$. If $|Var(\delta)|=3$, then $\delta=(y \vee y \vee z \vee z \vee w \vee w)$.
$C$ can be downgraded to a $C^2$ clause with $(x \vee y \vee z \vee w)$. 
If $|Var(\delta)|=4$, then we can either have $\delta=(y \vee y \vee z \vee z \vee w \vee u)$ or
$\delta=(y \vee y \vee z \vee z \vee w \vee w \vee u)$. We have $w=u$ for the former and
$u=0$ for the latter. If $|Var(\delta)|=5$, then (i) $\delta=(y \vee y \vee z \vee u \vee w \vee v)$  or
(ii) $\delta=(y \vee y \vee z \vee z \vee u \vee w \vee v)$  or 
(iii) $\delta=(y \vee y \vee z \vee z \vee u \vee u \vee w \vee v)$  or 
(iv) $\delta=(y \vee y \vee z \vee z \vee u \vee u \vee w \vee w \vee v)$ . 
For (iii), we can set $w=v$. For (iv), we can set $v=0$. For (i) and (ii), we branch $x=y=1$,
$x=\neg y$ and $x=y=0$. When $x=y=1$, we get a change of measure of $6$. 
When $x=\neg y$, then the remaining clause drops to a $C^2$ clause, this gives us 
$2+4\times(1-0.8376)=2.6496$. Finally, when $x=y=0$, we get a change of measure of $2$.
This gives us a branching factor of $\tau(6,2.6496,2)=1.4267$.
Finally, when $|Var(\delta)|\geq6$, then we branch $x=1$ and $x=0$. 
When $x=1$, $\delta$ drops to a $C^2$ clause, this gives us 
$1+6\times(1-0.8376)=1.9744$. On the other hand, when $x=0$, we have a change of measure of $1$.
This gives us a branching factor of $\tau(1.9744,1)=1.6237$.
\end{itemize}

This completes the case where a literal appears more than once in a $C^4$
clause. From this point onwards, all variables have weight 1. Hence, there will not be any 
increase in variable weights. \\

\item Let $C$ be a $C^4$ clause. 
At this point in time, all literals must only appear once in $C$.

\begin{itemize}
\item $|C|=4$. Assign all literals to $1$.
\item $|C|=5$. Let $C=(x_1 \vee x_2 \vee ... \vee x_5)$. Then we can negate all the literals
in $C$ to get $(\neg x_1 \vee \neg x_2 \vee ... \vee \neg x_5)$, and then $C$
becomes a $C^1$ clause and the overall measure drops. 
\item $|C|=6$. Let $C=(x_1 \vee x_2 \vee ... \vee x_6)$. Then we can negate all the literals
in $C$ to get $(\neg x_1 \vee \neg x_2 \vee ... \vee \neg x_6)$, and then $C$
becomes a $C^2$ clause and the overall measure drops. 
\item $|C|=7$. Let $C=(x_1 \vee x_2 \vee ... \vee x_7)$. hen we can negate all the literals
in $C$ to get $(\neg x_1 \vee \neg x_2 \vee ... \vee \neg x_7)$, and then $C$
becomes a $C^3$ clause and the overall measure drops.  

\item $|C|=8$. Let $C=(x \vee \delta)$. Now we branch $x=1$ and $x=0$.
When $x=1$, $\delta$ drops to a $C^3$ clause, this gives us 
$1+7\times(1-0.9412)=1.4116$. On the other hand, when $x=0$,
$\delta$ becomes a $C^4$ 7-literal clause, which can be downgraded to a $C^3$ clause,
giving us $1+7\times(1-0.9412)=1.4116$. Therefore, we have a branching
factor of $\tau(1.4116,1.4116)=1.6341$.  
\item $|C|=9$. Let $C=(x \vee y \vee \delta)$. Then we branch $x=y=1$ (first branch),
$x=\neg y$ (second branch) and $x=y=0$ (third branch). When $x=y=1$, we remove $x,y$, and $\delta$
drops to a $C^2$ 7-literal clause, giving us $2+7\times(1-0.8376)=3.1368$.
When $x=\neg y$, we remove $x$ and $\delta$ drops to a $C^3$ 7-literal clause,
giving us $1+7\times(1-0.9412)=1.4116$. Finally, when $x=y=0$, we have $\delta$
becomes a $C^4$ 7-literal clause, where we can further drop it to a $C^3$ 7-literal clause,
giving us $2+7\times(1-0.9412)=2.4116$. Therefore, we have a branching factor of 
$\tau(3.1368,1.4116,2.4116)=1.6500$. 

\item $|C|=10$. Let $C=(x \vee y \vee \delta)$. We again branch $x=y=1$ (first branch), 
$x=\neg y$ (second branch) and $x=y=0$ (third branch). 
Now when $x=y=1$, we remove $x,y$ and $\delta$ drops to a $C^2$ 8-literal clause,
giving us $2+8\times(1-0.8376)=3.2992$. When $x=\neg y$, we remove $x$ and $\delta$
drops to a $C^3$ 8-literal clause, giving us $1+8\times(1-0.9412)=1.4704$. When $x=y=0$, we remove $x,y$,
giving us a change of measure of $2$ and $\delta$ is a $C^4$ 8-literal clause. 
This gives us $\tau(3.2992,1.4704,2)$. For the first branch, $\delta$ is a $C^2$ 8-literal clause with a branching factor of 
$\tau(8\times0.8376,0.8376+6\times(0.8376-0.6464)-(1-0.8376),2\times0.8376)=\tau(6.7008,1.8224,1.6752)$.
For the third branch, we have a branching factor of $\tau(1.4116,1.4116)$. 
Applying vector addition to the first and third, we have the branching factor of 
$\tau(3.2992+6.7008,3.2992+1.8224,3.2992+1.6752,1.4704,2+1.4116,2+1.4116)=1.6545$.

\item $|C|=11$. Let $C=(x \vee y \vee \delta)$. We branch $x=y=1$ (first branch), $x=\neg y$ (second branch)
 and $x=y=0$ (third branch). When $x=y=1$, $\delta$ drops to a $C^2$ 9-literal clause, 
giving us a change of measure of $2+9\times(1-0.8376)=3.4616$. When $x=\neg y$, we remove $x$
and $\delta$ drops to a $C^3$ 9-literal clause, giving us a change of measure of 
$1+9\times(1-0.9412)=1.5292$. Finally, when $x=y=0$, we have a change of measure of $2$.
This gives us an initial branching factor of $\tau(3.4616,1.5292,2)$. For the first branch, we have a $C^2$ 9-literal clause,
with a branching factor of 
$\tau(9\times0.8376,0.8376+7\times(0.8376-0.6464)-(1-0.8376),2\times0.8376)=\tau(7.5384,2.0136,1.6752)$.
Applying vector addition to the first branch, we get a branching factor of 
$\tau(3.4616+7.5384,3.4616+2.0136,3.4616+1.6752,1.5292,2)=1.6335$.

\item $|C|\geq12$. Let $C=(x \vee y \vee \delta)$. We branch $x=y=1$, $x=\neg y$, $x=y=0$.
When $x=y=1$, $\delta$ drops to at least a $C^2$ 10-literal clause, 
giving us a change of measure of $2+10\times(1-0.8376)=3.624$. When $x=\neg y$, we remove $x$
and $\delta$ drops to at least a $C^3$ 10-literal clause, giving us a change of measure of 
$1+10\times(1-0.9412)=1.588$. Finally, when $x=y=0$, we have a change of measure of $2$.
This gives us a branching factor of at most $\tau(3.624,1.588,2)=1.6354$.
\end{itemize}
\end{enumerate}

\noindent
We have therefore covered all the cases needed in the algorithm. 
This proves the correctness of the algorithm as well. By comparing all the branching factors obtained
above, we have the following result:

\begin{theorem}
The algorithm that solves G$4$XSAT runs in $O(1.6545^n)$ time.
\end{theorem}

\section{Exponential Space Algorithms}

\noindent
Let $m$ denote the number of clauses in the formula $\varphi$.
Here, we present faster exact algorithms to solve G$i$XSAT. This is a
modification of Schroeppel and Shamir's result to bring down
the complexity of generalised XSAT and Knapsack algorithms from
$O(2^n)$ to $O(2^{n/2})$ using exponential space.
Roughly speaking, they just note down for each clause how many
satisfying literals are required and then they calculate for each
possible assignment of the first half of variables how many of
literals of each clause are satisfied. Each of these
up to $2^{n/2}$ vectors is written into a database; note that by standard
database techniques, reading (= checking whether a vector
is in the database) and writing into a database
takes time logarithmic in the size of the database and
thus is polynomial time for our purposes.
After that one processes the variables of the second half
and for each possible assignment $\bf y$, one computes the vector
$\bf s$ of literals which need to made true in the first half
of the variables in order to have a match; if this vector
is in the database then a solution exists. One major achievement
of Schroeppel and Shamir \cite{SS81} is that their space usage
is roughly $O(2^{n/4})$ instead of $O(2^{n/2})$ what our sketch
of the algorithm gives; we do not intend to optimise the space here and
do therefore settle for the conceptually easier algorithm outlined.
Dahll\"of verified that the result of Schroeppel and Shamir is
indeed applicable to G$i$XSAT and solved this this problem for
$i=2,3,4$ in $O(1.4143^n)$ time.

In this section, we modify the results of Schroeppel and Shamir
to derive faster exact algorithms to solve G$i$XSAT in exponential space.
We do this with an asymmetric splitting in which we exploit that this
splitting of the variables can be done such that for the larger half of
the variables, the vectors are all the solutions of the smaller G$i$XSAT
problem which consists of all clauses fully covered by the half they are
in and as the larger half is a union of clauses,
one therefore can enumerate all the possible solutions
of that smaller problem only which will be less than $2$ to the number
of variables; these savings then allows
to handle $\alpha n$ variables instead of
$0.5 n$ variables for some larger $\alpha$ and the number of tuples
generated and time used is then actually equal within $O(2^{(1-\alpha)n})$.
For the smaller half, one computes for every possible choice of the
remaining variables the corresponding vector to be checked to be
in the data base for a match and if found, the instance of the
G$i$XSAT problem is satisfiable. We now explain our algorithm and
then the verification more formally.

\medskip
\noindent
Algorithm to solve G$i$XSAT \\
Input : $\varphi$. \\
Output : If we find with the below algorithm that $\varphi$ is solvable
then we output $1$ else we output $0$ (see Step 5 of the algorithm).

\begin{enumerate}
\spaceo
\item We choose $\alpha$ strictly between $0$ and $1$.
Then we split the $n$ variables into
two groups, one containing $\alpha n$ many, while the other containing
$(1-\alpha)n$.
For the $\alpha n$ variables, we choose them such that they are all those
variables which belong to some set of clauses $S$; in order to meet our
cut-off, there can be one clause which is used only partially, say it has
$h+k$ variables which $h$ going into the set and $k$ not being in. Then
we split it into a $h$-literal and $k$-literal clause and branch over
the $i+1$ cases on how many of the $h$ literals are true in the clause
of $S$; this only imposes a constant additional factor on the runtime
of the algorithm. Let $S'$ be the set of all other clauses.
\item Branch clause by clause for each clause $C \in S$; for this we
      substitute the fixed variables in a branched clause in all other
      clauses by their values so that some $C^j$ clauses can be reduced
      to $C^\ell$ clauses for some $\ell < j$ where $j-\ell$ is the number
      of literals in the clause already set to $1$; in some cases like
      $\ell$ being negative, the corresponding branch is unsatisfiable.
\item Note down for each possible choice of the $\alpha n$ variables and
      for each clause in $S'$, how many literals in each of these clauses
      they made true (and discard solutions which satisfy too many or too
      few literals) and write the corresponding vector into a 
      database (which can be updated and checked in polynomial time).
\item For every possible choice of values of the remaining $(1-\alpha) n$ variables, 
       compute the vector of how many literals in each clause in S' must have been 
       satisfied by the other half of the splitting to combine to a full solution 
       and check whether this vector is in the database.
\item If for one possibility the corresponding vector is found in the database
      then output $1$, else output $0$.
\end{enumerate}

\subsection{Single Occurrence}

\noindent
We first investigate the case that each variable occurs in a clause at most
once. This is a restriction assumption, but it is easier to deal with
mathematically. Once we have obtained the values for the single occurrence
case, we will show that the values for the multi-occurrence case (where
a variable can occur several times) are the same.
Note that the choosing the value of $\alpha$ is different for each G$i$XSAT,
$i>1$. When branching clause by clause, we do not take all 
combinations of the variables, but only those which make the clause true.
This allows us to improve the time further.

We give here the values of $\alpha$ for G$2$XSAT, G$3$XSAT and G$4$XSAT.
The numbers in the table are ${k \choose h}^{1/k}$. This are the branching
factor, if one branches a $C^h$ clause with all $k$ distinct variables
(independent on whether these are negated). Note that only $C^h$ clauses
with $h \leq k/2$ are interesting, as one can negate all variables in a
$C^h$ clause to obtain an equivalent $C^{k-h}$ clause. Furthermore,
if one branches only the first variable in a repetition-free
$C^h$ clause with $k$ variables and if $k \leq k/2$
then one either has to solve a
$C^{h-1}$ clause of $k-1$ variables or a $C^h$ clause with $k-1$ variables.
As ${k \choose h} = {k-1 \choose h-1}+{k-1 \choose h}$ and as branching
a ${k \choose 1}$ clause has $k$ possibilities (depending on which literal
is $1$), one has for the problem to determine the branching factor of
repetition-free $C^h$ $k$-literal clauses that this one is ${k \choose h}$
possibilities for $k$ variables; now in order to normalise this to the average
of $1$ variable, the branching factor is ${k \choose h}^{1/k}$ per variable.
This allows to compute the overall branching factor, when brute-forcing
clause by clause (and not variable by variable) a repetition-free
G$i$XSAT instant of $C^h$ clauses with $h \leq i$ and $h \leq k/2$
(where $k$ is the number of variables in the current clause considered)
is then the largest number $c$ occuring within the first $i$ column in of
the corresponding entry in the first or second table below (where void
entries are ignored as there $h > k/2$). If one does
the splitting of variables in $\alpha n$ variables branched clause by clause
and $(1-\alpha) n$ variables branched by brute force per variable,
then one needs that $q^\alpha = 2^{1-\alpha}$ for the optimal $\alpha$
chosen. The $k$ in these calculations refers also to the row number in the
table.

\subsection{G2XSAT}

\begin{center}
\begin{tabular}{ |c|c|l| } 
 \hline
 Length of clause & Time Complexity & Branching factor for $h=1,2$ w/o $\alpha$ \\
\hline
 $k$ & formula for ${k \choose h}^{\alpha n/k}$ & value ${k \choose h}^{1/k}$\\
\hline
 2 & $\binom{2}{h}^{\fracy{\alpha n}{2}}$  & 1.4143 \\ 
 3 & $\binom{3}{h}^{\fracy{\alpha n}{3}}$  & 1.4423, 1.4423 \\ 
 4 & $\binom{4}{h}^{\fracy{\alpha n}{4}}$  & 1.4143, 1.5651 \\ 
 5 & $\binom{5}{h}^{\fracy{\alpha n}{5}}$  & 1.3798, 1.5849 \\
 6 & $\binom{6}{h}^{\fracy{\alpha n}{6}}$  & 1.3481, 1.5705 \\ 
 7 & $\binom{7}{h}^{\fracy{\alpha n}{7}}$  & 1.3205, 1.5449 \\ 
 \hline
\end{tabular}
\end{center}

\noindent
In the table above, the values in each row denote the branching factor of a clause of
that length. The first value denotes the branching factor of branching a $C^1$ clause,
while the second value denotes the branching factor of branching a $C^2$ clause. For example,
for the clause of length $5$, we have $\binom{5}{1}^{1/5} = 5^{1/5} = 1.3798$ 
and $\binom{5}{2}^{1/5} = 10^{1/5} = 1.5849$.

Let $k$ denote the length of the clause. 
For the verification that for $k \geq 7$ it holds that
$(k(k-1)/2)^{1/k} \leq 1.58$, note that one can write this
as $e^{(\log(k)+\log(k-1)-\log(2))/k} \leq 1.58$, where
$e$ is Euler's constant of approximately 2.71828 and
$\log$ is in this paragraph the natural logarithm. Now this is equivalent
to $\log(k)+\log(k-1)-\log(2) \leq \log(1.58) \cdot k$
and $\log(k)+\log(k-1)-\log(2)-\log(1.58) \cdot k \leq 0$.
The derivative of this is
$1/k+1/(k-1)-\log(1.58)$ and as $\log(1.58)$ is approximately $0.4574$,
for $k \geq 6$, $1/k+1/(k-1) \leq 0.4$ and the derivative is negative.
So for all $k \geq 6$ it holds that
$\log(k)+\log(k-1)-\log(2)-\log(1.58) \cdot k$ does not grow and has
a negative value, as the term is at $k=6$ already approximately $-0.0365$.
This justifies the conclusion, that $1.5849$
is the maximal value of a term $(k \cdot (k-1) / 2)^{1/k}$ for all $k$
which are natural numbers; not only for those $k$ in the table but
also for the larger ones. Similarly one can derive that
$k^{1/k}$ is bounded by $1.5849$ for all natural numbers $k$.

From the table, we see the max branching factor comes from branching
$C^2$ 5-literal clauses, which gives $1.5849$. 

Since we want the overall timing for both sides of the split to
be the same, we solve for $\alpha$ by
having $1.5849^{\alpha n} = 2^{(1-\alpha)n}$. Therefore, this gives
$\alpha = 0.600823$. Since $1.5849^{0.600823n}=1.3188^n$, 
our G$2$XSAT algorithm runs in $O(1.3188^n)$ time.

\subsection{G3XSAT and G4XSAT}

\noindent
We expand the number of columns and rows
in the table from the previous section.

\begin{center}
\begin{tabular}{ |r|c|l| } 
 \hline
 Length of clause & Time Complexity & Branching factor for $h=1,2,3,4$ w/o $\alpha$ \\
\hline
 $k$ & formula for ${k \choose h}^{\alpha n/k}$ & value ${k \choose h}^{1/k}$\\
\hline
 2 &   $\binom{2}{h}^{\fracy{\alpha n}{2}}$  & 1.4143 \\ 
 3 &   $\binom{3}{h}^{\fracy{\alpha n}{3}}$  & 1.4423, 1.4423 \\ 
 4 &   $\binom{4}{h}^{\fracy{\alpha n}{4}}$  & 1.4143, 1.5651 \\ 
 5 &   $\binom{5}{h}^{\fracy{\alpha n}{5}}$  & 1.3798, 1.5849, 1.5849 \\
 6 &   $\binom{6}{h}^{\fracy{\alpha n}{6}}$  & 1.3481, 1.5705, 1.6476 \\ 
 7 &   $\binom{7}{h}^{\fracy{\alpha n}{7}}$  & 1.3205, 1.5449, 1.6619, 1.6619 \\ 
 8 &   $\binom{8}{h}^{\fracy{\alpha n}{8}}$  & 1.2969, 1.5167, 1.6540, 1.7008 \\ 
 9 &   $\binom{9}{h}^{\fracy{\alpha n}{9}}$  & 1.2766, 1.4891, 1.6361, 1.7115 \\ 
10 & $\binom{10}{h}^{\fracy{\alpha n}{10}}$  & 1.2590, 1.4633, 1.6141, 1.7070 \\ 
11 & $\binom{11}{h}^{\fracy{\alpha n}{11}}$  & 1.2436, 1.4396, 1.5908, 1.6942 \\ 
 ... & ...  & ... \\   
 \hline
\end{tabular}
\end{center}

\noindent 
We expand the number of columns and rows
in the table from the previous section. The values are interpreted in the similar manner
as in the previous table.

For G$3$XSAT,
from the above table, we see that the max branching factor comes
from branching $C^3$ 7-literal clauses, which gives $1.6619$.
Again, we solve for $\alpha$ by having $1.6619^{\alpha n}=2^{(1-\alpha)n}$.
This gives $\alpha=0.57712$. Therefore, $1.6619^{0.57712n}=1.3407^n$
and G$3$XSAT algorithm runs in $O(1.3407^n)$ time.

For G$4$XSAT, we see from the same table that the max branching
factor comes from $C^4$ 9-literal clauses, which gives $1.7115$.

Taking $1.7115^{\alpha n}=2^{(1-\alpha)n}$, we have 
$\alpha=0.5633$. Therefore $1.7115^{0.5633n}=1.3536^n$.
Hence, our G$4$XSAT algorithm takes $O(1.3536^n)$ time.

\subsection{Multi-occurrences}
Note that in the previous section, for our G$i$XSAT algorithm,
$i \in \{2,3,4\}$, we made an assumption that each literal appearing
in any clause only appears once. However, it is possible that a literal
in that clause may appear more than once. We assume that a $C^j$ clause
of the form $x \vee \neg x \vee \alpha$ is simplified to the $C^{j-1}$ clause
$\alpha$ and therefore we assume that every literal either occurs
only positive or only negative. To keep notation simple, as we treat clause
by clause, we assume that in the clause investigated, all variables appear
positive.

In this section, we show that even if such cases happen, the number of cases
that we can have to satisfy our clauses are satisfied are bounded above by
the number of cases that we have shown in the previous section. This
therefore implies that we get a better branching factor. Hence, the worst
case is such that each literal appears exactly once in any clause. 

Given a $C^j$ clause $C$, for $j  \in \{1,2,3,4\}$, we let $F_j(\ell)$
denote the number of possible ways to satisfy that clause
where $|Var(C)|=\ell$. 
Then we define $F(\ell,h)$ as the maximum number of ways
to satisfy a clause $C^j$ clause, with $j\leq h$ which has $\ell$ variables.
If $\ell=0$, then $F(\ell,h) = 1$, for any $h$. Next, we define
$G(\ell,h)$, which is the maximum number of possible ways to satisfy a $C^j$
clause, $j\leq h$, with
$\ell$ many variables, where each variable appears only once in that clause;
$G(\ell,h)$ are therefore the numbers which we investigated in the
single-occurrence case. By
definition, we have $G(\ell,h) = max\{ \binom{\ell}{j} : j\leq h\}$.
We recall that in the sections for the single occurrence case
and the tables there we used that
$G(\ell,h) = \binom{\ell}{h}$ in the case that $\ell \geq 2h-1$ and
$G(\ell,h) = \binom{\ell}{k}$ in the case that one can
choose this $k$ such that $k \leq h$ and $2k-1 \leq \ell \leq 2k+1$.

Our goal is now to show that $F(\ell,h) \leq G(\ell,h)$
for all $\ell$ and $h \in \{1,2,3,4\}$. For $\ell \leq8$, we
run a computer program to display the values of $F(\ell,h)$ and $G(\ell,h)$
to show that $F(\ell,h) \leq G(\ell,h)$. After which, we prove by
induction that $F(\ell,h) \leq G(\ell,h)$, for $\ell > 8$,
for $h\in \{1,2,3,4\}$.
\begin{enumerate}
\spaceo
\item $h=1$. Then $F(\ell,1) = \binom{\ell}{1}$.
Every literal in $C$ must appear
exactly once. Therefore, we have $\ell$ many possible
options to satisfy $C$. 
Similarly we have $G(\ell,1) = \binom{\ell}{1}$.
Therefore, $F(\ell,1) \leq G(\ell,1)$.

\item $h=2$. Let $b$ denote the number of variables that appears
twice (as the same literal) in $C$ and $a$ denote the number of variables
that appear once in $C$. Note that $a+b=\ell$. Then $F_2(\ell,a,b)$ 
denotes the number of possible combinations to satisfy a $C^2$ clause
given that it has $\ell$ number of variables, with $a$ many variables
appearing once
and $b$ many variables appearing twice. We define $F(\ell,2)$ as the
maximum number of possible choices to satisfy a $C^j$ clause, $j\leq 2$,
with $\ell$ many number of variables. Hence 
$F(\ell,2) = max\{F_2(\ell,a,b),F(\ell,1) : a+b = \ell \}$.
We give the recursion formula for $F_2(\ell,a,b)$ below.

\begin{itemize}
\item $F_2(0,0,0)=0$.
When there are no variables in this clause, $C$ is unsatisfiable.
\item $F_2(1,1,0)=0$.
When there are only 1 variable in $C$, with only 1 variable appearing once,
then $C$ is unsatisfiable.
\item $F_2(1,0,1)=1$.
When there is a variable appearing twice in a $C^2$ clause, then
there is only 1 way to satisfy this clause, by assigning the literals to $1$. 
\item For $\ell \geq 2$, we define the following :
\[
F_2(\ell,a,b)= 
\begin{cases}
    F_2(\ell-1,a,b-1) + 1 ,& \text{if $b>0$;} \\
    \binom{a}{2}, & \text{if $b=0$.}
\end{cases}
\]
The recursion formula of $F_2(\ell,a,b)$
can be seen as follows : We choose any variable
$x$ appearing twice to branch. If $x=1$,
then the clause is satisfied and this counts as
a way of satisfying the clause. Hence the ``+1'' in
the formula. On the other hand, when $x=0$,
then the number of choices to satisfy it is given
in $F_2(\ell-1,a,b-1)$. When $b=0$, then
we are only left with variables appearing once in the formula.
Then to satisfy the $C^2$ clause,
there are $\binom{a}{2}$ many choices to satisfy it. 

With this, we compare $F(\ell,2)$ with $G(\ell,2)$ below :

\begin{center}
\begin{tabular}{|r|r|r|r|r|r|r|r|r|}
\hline
$\ell$ & $1$ & $2$ & $3$ & $4$ & $5$ & $6$ & $7$ & $8$ \\
\hline
$F(\ell,2)$ & $1$ & $2$ & $3$ & $6$ & $10$ & $15$ & $21$ & $28$\\
\hline
$G(\ell,2)$ & $1$ & $2$ & $3$ & $6$ & $10$ & $15$ & $21$ & $28$ \\
\hline
\end{tabular}
\end{center}

\smallskip
\noindent
With this, we see that $F(\ell,2) \leq G(\ell,2)$ for all $\ell \leq 8$.
\end{itemize}

\item $h=3$. Let $c$ denote the number of variables that appears
three times in $C$, $b$ denote the number of variables that appear twice 
and $a$ denote the number of variables that appear once in $C$. Note that 
$a+b+c=\ell$. Then $F_3(\ell,a,b,c)$ denotes the number of possible 
combinations to satisfy a $C^3$ clause given that
it has $\ell$ number of variables, with $a$ many variables appearing once
, $b$ many variables appearing twice and $c$ many variables appearing
three times. We define $F(\ell,3)$ as the
maximum number of possible choices to satisfy a $C^j$ clause, $j\leq 3$,
with $\ell$ many number of variables. Hence 
$F(\ell,3) = max\{F_3(\ell,a,b,c),F(\ell,2) : a+b+c = \ell \}$
($F(\ell,2)$ was defined in the case $h=2$). 
We give the recursion formula for $F_3(\ell,a,b,c)$ below.

\begin{itemize}

\item $F_3(0,0,0,0)=0$.
If there are no variables, then this clause is not satisfiable.
\item $F_3(1,1,0,0)=0$.
If there is only 1 variable that also appears once,
then this clause is not satisfiable.
\item $F_3(1,0,1,0)=0$.
If there is only 1 variable that also appears twice,
then this clause is not satisfiable.
\item $F_3(1,0,0,1)=1$.
If there is only 1 variable that also appears three times, 
then the only way to satisfy this clause is to assign this literal as 1. 

\item For $\ell \geq 2$, we define the following :
\[
F_3(\ell,a,b,c)= 
\begin{cases}
    F_3(\ell-1,a,b,c-1) + 1 ,& \text{if $c>0$;} \\
    F_3(\ell-1,a,b-1,c) + a, & \text{if $c=0$ and $b>0$;}\\
    \binom{a}{3} ,& \text{if $c=0$ and $b=0$} 
\end{cases}
\]
The recursion formula of $F_3(\ell,a,b,c)$ can be seen as follows :
We choose any variable
$x$ appearing three times to branch. If $x=1$, then the clause
is satisfied and this counts as
a way of satisfying the clause. Hence a ``+1'' in the formula.
On the other hand, when $x=0$,
then the number of choices to satisfy it is given in $F_3(\ell-1,a,b,c-1)$.
When $c=0$, then we are only left with variables appearing at most twice
in the formula. Suppose now we choose any
variable $y$ appearing twice to branch.
When $y=1$, then the clause becomes a $C^1$ clause
where we can choose any of the $a$ variables
to follow up to branch. This explains the ``+a'' portion. 
When $y=0$, we have $F_3(\ell-1,a,b-1,c)$
many choices left. Finally, when $b=c=0$, then to satisfy
a $C^3$ clause, we can have $\binom{a}{3}$
choices to choose from. This explains the recursion formula.

With this, we compare $F(\ell,3)$ with $G(\ell,3)$ below : \\ \mbox{ }
\begin{center}
\begin{tabular}{|r|r|r|r|r|r|r|r|r|}
\hline
$\ell$ & $1$ & $2$ & $3$ & $4$ & $5$ & $6$ & $7$ & $8$ \\
\hline
$F(\ell,3)$ & $1$ & $2$ & $3$ & $6$ & $10$ & $20$ & $35$ & $56$ \\
\hline
$G(\ell,3)$ & $1$ & $2$ & $3$ & $6$ & $10$ & $20$ & $35$ & $56$ \\
\hline
\end{tabular}
\end{center}

\smallskip
\noindent
With this, we see that $F(\ell,3) \leq G(\ell,3)$ for all $\ell \leq 8$.
\end{itemize}

\item $h=4$. Let $d$ denote the number of variables that appears
four times in $C$, $c$ denote the number of variables that appears three times
, the number of variables that appear twice as $b$ and finally the number of 
variables that appear once as $a$. Note that $a+b+c+d=\ell$. 
Then $F_4(\ell,a,b,c,d)$ denotes the number of possible 
combinations to satisfy a $C^4$ clause given that
it has $\ell$ number of variables, with $a$ many variables appearing once
, $b$ many variables appearing twice, $c$ many variables appearing
three times and finally $d$ many variables appearing four times. 
We define $F(\ell,4)$ as the maximum number of possible choices to satisfy a $C^j$ clause, $j\leq 4$, 
with $\ell$ many number of variables. Hence  
$F(\ell,4) = max\{F_4(\ell,a,b,c,d),F(\ell,3) : a+b+c+d= \ell \}$ ($F(\ell,3)$ was defined in the case $h=3$). 
We give the recursion formula for $F_4(\ell,a,b,c,d)$ below.
\begin{itemize}
\item $F_4(0,0,0,0,0)=0$. If there is a clause with no variables, then it is not satisfiable.
\item $F_4(1,1,0,0,0)=0$. If there is a variable appearing once, then it cannot satisfy a $C^4$ clause.
\item $F_4(1,0,1,0,0)=0$. If there is a variable appearing twice, then it cannot satisfy a $C^4$ clause.
\item $F_4(1,0,0,1,0)=0$. If there is a variable appearing three times, then it cannot satisfy a $C^4$ clause.
\item $F_4(1,0,0,0,1)=1$. If there is a variable appearing four times, then that literal must be assigned 1,
which is the only way to satisfy this clause.

\item For $\ell \geq 2$, we define the following :
\[
F_4(\ell,a,b,c,d)= 
\begin{cases}
    F_4(\ell-1,a,b,c,d-1) + 1 ,& \text{if $d>0$;} \\
    F_4(\ell-1,a,b,c-1,d) + a, & \text{if $d=0$ and $c>0$;}\\
    F_4(\ell-1,a,b-1,c,d) + (b-1)+\binom{a}{2}, & \text{if $c=d=0$ and $b>0$;}\\
    \binom{a}{4} ,& \text{if $b=c=d=0$.} 
\end{cases}
\]
\noindent
The recursion formula of $F_4(\ell,a,b,c,d)$ can be seen as follows : We choose any variable
$x$ appearing four times to branch. If $x=1$, then the clause is satisfied and this counts as
a way of satisfying the clause. Hence a ``+1'' in the formula. On the other hand, when $x=0$,
then the number of choices to satisfy it is given in $F_4(\ell-1,a,b,c,d-1)$. When $d=0$, then
we are only left with variables appearing at most three times in the clause. Suppose now we choose any
variable $y$ appearing three times to branch. When $y=1$, then the clause becomes a $C^1$ clause
where we can choose any $a$ many variable to follow up to branch. This explains the ``+a'' portion. 
When $y=0$, we have $F_4(\ell-1,a,b,c-1,d)$ many choices left. Next, if $c=0$, then the clause
contains only variables that appear at most twice. We choose any variable $z$ that appears twice in the clause 
to branch. When $z=1$, the clause becomes a $C^2$ clause. Here, we can choose the from the 
remaining $b-1$ variables (that appear twice) or choose any two of the variables that appear once
($\binom{a}{2}$). Therefore, we have $(b-1)$+$\binom{a}{2}$. On the other hand, when $z=0$,
we have $F_4(\ell-1,a,b-1,c,d)$ many choices. Finally, when $b=0$, then to satisfy
a $C^4$ clause, we can have $\binom{a}{4}$ choices to choose from. This explains the recursion formula. 

With this, we compare $F(\ell,4)$ with $G(\ell,4)$ below :

\begin{center}
\begin{tabular}{|r|r|r|r|r|r|r|r|r|}
\hline
$\ell$ & $1$ & $2$ & $3$ & $4$ & $5$ & $6$ & $7$ & $8$ \\
\hline
$F(\ell,4)$ & $1$ & $2$ & $3$ & $6$ & $10$ & $20$ & $35$ & $70$ \\
\hline
$G(\ell,4)$ & $1$ & $2$ & $3$ & $6$ & $10$ & $20$ & $35$ & $70$ \\
\hline
\end{tabular}
\end{center}

\noindent
\smallskip
With this, we see that $F(\ell,4) \leq G(\ell,4)$, $\ell \leq 8$.
\end{itemize}
\end{enumerate}

\noindent
Now, for any $h \in \{1,2,3,4\}$, we prove the remaining $\ell > 8$
by induction,
with $\ell \leq 8$ shown above. Suppose by induction hypothesis that
$F(\ell,h) \leq G(\ell,h)$. Now for all $\ell>8$, we know that
$\binom{\ell}{1} < \binom{\ell}{2} < \binom{\ell}{3} < \binom{\ell}{4}$.
Therefore, $G(\ell,h) = \binom{\ell}{h}$. We need to show that 
$F(\ell + 1,h) \leq G(\ell + 1,h)$.

Let $x$ be a variable in a $C^j$ clause, $j\leq h$, with $\ell+1$
many variables. When $x=0$, then there are at most $F(\ell,h)$
many possibilities
to satisfy the clause. When $x=1$, $C^j$ clause can be satisfied, or drop to a
$C^{j'}$ clause, for $j'<j$. Therefore we can have at most $F(n,j')$ many
possible ways to satisfy the clause when $x=1$. 
Therefore $F(\ell+1,h) \leq max \{ F(\ell,h)+F(\ell,j) : j < h \}$.
Note that for any
$\ell$, we have $F(\ell,1) \leq F(\ell,2) \leq F(\ell,3) \leq F(\ell,4)$.
Therefore,
$F(\ell+1,k) \leq F(\ell,k) + F(\ell,k-1) \leq G(\ell,k) + G(\ell,k-1) = 
\binom{\ell}{k} + \binom{\ell}{k-1} = \binom{\ell+1}{k} \leq G(\ell+1,k)$.
This completes the inductive proof. Therefore, even if multi-occurrences
occur in a clause, the worst case scenario of the algorithm still comes
from dealing with clauses with every single variable appearing once
in the clause.

\section{Summary and Future Work}

\noindent
In this paper, we gave polynomial space algorithms to solve G$2$XSAT
in $O(1.3674^n)$ time, G$3$XSAT in $O(1.5687^n)$ time and G$4$XSAT
in $O(1.6545^n)$ time. These algorithms 
were designed with a nonstandard measure in mind
and the use of state based measure to optimize
the current time bound for our G$2$XSAT algorithm. 
For our G$3$XSAT and G$4$XSAT algorithms, our goal is to introduce
simple algorithms (removing $C^1$, $C^2$, $C^3$ clauses in that order)
with the help of nonstandard measure to improve the current state
of the art algorithm, without having to deal with
an explosion of cases. However, we did not spend as much effort on
these algorithms as we did for the G$2$XSAT algorithm, so there might
still be some room to improve the bound further.

We also gave exponential space algorithms to solve
G$2$XSAT, G$3$XSAT and G$4$XSAT.
The idea here is not to split the variables asymmetrically,
on one side, branching clause after
clause, while we brute force the variables on the other side.
This allows us to improve on the
result gave by Schroeppel and Shamir $(O(1.4143^n))$ where our bounds
are then $O(1.3188^n)$ for G$2$XSAT, $O(1.3407^n)$ for G$3$XSAT and
$O(1.3536^n)$ for G$4$XSAT. We note that this can be done for all
G$i$XSAT with $i \geq 2$ and that the sequence of the corresponding
complexities converges from below to $O(\sqrt{2}^n)$.

Furthermore, our polynomial space algorithms are quite complex.
Future investigations might try to find better ways to improve the
algorithm timings without getting the explosion of cases which every
small improvement on the current timings gave in our attempts; however,
we ourself did not find a way to do this.

\end{document}